\documentclass[12pt]{article}  %%%%the type of document you're going to make

\usepackage{graphicx} 
\usepackage{amsmath}
\usepackage{amsfonts}

\newcommand{\commentout}[1]{}

\usepackage[left=2cm,top=2cm,right=2cm,bottom=2cm,nohead,nofoot]{geometry}
\usepackage{setspace}
\usepackage{natbib}
\usepackage[usenames,dvipsnames]{color}

\newcommand{\imloc}{}%images/}

\author{A.C. Thomas, Samuel L. Ventura, Shane Jensen, Stephen Ma}
\title{Competing Process Hazard Function Models for Player Ratings in Ice Hockey}  %Hierarchical Bayesian 
\date{\today}

\begin{document}
\maketitle{}
\begin{abstract}

Evaluating the overall ability of players in the National Hockey League (NHL) is a difficult task. Existing methods such as the famous ``plus/minus" statistic have many shortcomings. Standard linear regression methods work well when player substitutions are relatively uncommon and scoring events are relatively common, such as in basketball, but as neither of these conditions exists for hockey, we use an approach that embraces the unique characteristics of the sport. We model the scoring rate for each team as its own semi-Markov process, with hazard functions for each process that depend on the players on the ice. This method yields offensive and defensive player ability ratings which take into account quality of teammates and opponents, the game situation, and other desired factors, that themselves have a meaningful interpretation in terms of game outcomes.  Additionally, since the number of parameters in this model can be quite large, we make use of two different shrinkage methods depending on the question of interest: full Bayesian hierarchical models that partially pool parameters according to player position, and penalized maximum likelihood estimation to select a smaller number of parameters that stand out as being substantially different from average. We apply the model to all five-on-five (full-strength) situations for games in five NHL seasons.
\end{abstract}

%\begin{keywords}
%Semi-Markov Processes, Hazard Functions, Lasso, Ratings, Hockey
%\end{keywords}

\newpage

\doublespacing
% Section 1: Introduction

\section{Introduction}

In many situations where a desired outcome depends on the performance of a group, it can be difficult to evaluate the individual contributions of its members. The study of sports provides a number of examples; the easier decomposition of baseball into what are essentially head-to-head match-ups makes it comparatively easy to tell whether one batter is superior to another, given enough observations.

The study of goal-based team sports -- ice hockey, field hockey, basketball, soccer, and lacrosse, among others -- is considerably more difficult, as the separation of roles is much more difficult to measure with modern game statistics, especially when player efforts do not directly lead to goals. In ice hockey, player abilities are historically quantified by citing offensive statistics, such as goals and assists, defensive statistics such as blocked shots and a goaltender's saves, and combinations such as the Plus/Minus statistic (+/-), or the net goals scored for a player's team when that player was on the ice. However, these are measured across many different combinations of players on the ice who contribute to the play, so an overall assessment of individual ability is not as obvious. Even if we assume that goaltenders have no role in team offense, there is surely a defensive assessment that can be made for other players, which is not as easily captured by these count-based statistical measures.

The nature of ice hockey means that scoring events are often quite rare. If we divide a game into many segments when the total number of goals scored is less than ten, the majority of these may be empty of scoring events, requiring a treatment that is considerate of this imbalance; segments of unequal length must also be handled appropriately. 

This rarity also contributes to another important consideration -- what if the data are insufficient to adequately separate players from each other in their ratings, or have little to no predictive value, either for a player's own future performance or for in-game outcomes? Any method we use to generate these ratings should take this into account, either as an integral part of the method or as a post-analysis check.

%We don't believe that every player's ability can be distinguished with this data, so we use variable selection to optimize the rating system such that the number of "exceptional" players is far smaller than the population -- some good, some bad, most average. 

To manage these factors and generate meaningful player ratings, we propose to measure the abilities of players in ice hockey according to goal-scoring rates when they are on the ice, much as in the plus/minus approach. However, we have two particular features of our approach that improve upon plus/minus. First, we consider goal-scoring to be the combination of at least two semi-Markov processes, modulated by the players on the ice for each team, so that each player on the ice contributes to both their team offense and team defense. Second, we regularize these estimates to ensure better predictive performance, which may also have the benefit of selecting a subset of players to have non-zero (i.e. non-average) ratings. %The Lasso method \citep{tibshirani1996rsasvl} has each of these properties; methods that derive from this may also be adapted to the semi-Markov setting.

Ideally, our method for obtaining meaningful player ratings will have several important properties. We want ratings that can be interpreted in terms of game outcomes -- namely, goals scored or prevented. In that spirit, we want to distinguish the offensive and defensive capabilities of each player separately, allowing for a superior assessment of ability, as well as the quality of a player's team, teammates and opposition by factoring their abilities into each observed event. In some cases, we would also like to distinguish a subset of players as ``exceptional'' at offense or defense (in either direction). 

We continue by describing previous methods for rating the offensive and defensive skill for players in hockey and other sports in Section \ref{s:previous-work}, as well as describing the data available for this work. In Section \ref{s:model-spec} we describe our methodological approach to the problem, demonstrating many of its applications in Sections \ref{s:applications} and \ref{s:extra-applications}. We conclude in Section \ref{s:discussion} by discussing potential extensions to our approach.   %; we verify that our methods and models are adequate, and compare them explicitly with others, in Section \ref{s:model-val}

\section{Previous Approaches for Player Ratings}\label{s:previous-work}

\subsection{Count-Based Measures: Simple Plus/Minus}

The notion of tracking the number of goals scored, both for and against, for each player on the ice is decades old, but its full application took years to reach its current state. In the National Hockey League (NHL), the world's premier professional ice hockey organization, its initial use  was said to be pioneered by the Montreal Canadiens hockey club in the 1950s, though only for their own purposes and in secret. The system was popularized by NHL coach Emile Francis during the 1960s, though the existing weaknesses of this approach were obvious even then: it does not take into account each player's quality of teammates, quality of opponents, and position. Good players on bad teams often have similar plus/minus statistics as bad players on good teams.  

Without changing the basic structure of the statistic, the most obvious weakness one can address is the effective rarity of goals, an average of roughly three per team per game. By adding other events that can lead to goals, more information can be attributed to the efforts of players on the ice. These typically include shots on goal, either unweighted or adjusted for the distance from the net, possibly including those that are blocked by the opposing team's skaters or miss the net entirely; these include the Fenwick- and Corsi- weighted Plus/Minus; \citet{macdonald2012egmfentap} lists these and others that have been adapted to the general approach.

\citet{lock2009b+rscnp} and \citet{schuckers2011nhlsrbuaoeampaa} extend this idea by accounting for all events that are recorded in a modern NHL game, including faceoffs, turnovers, and hits, all of which are thought to change the likelihood of the scoring of goals, either due to changes in puck possession or location on the ice. Each of these has an effective ``weight'' in terms of the expected number of goals scored or prevented because that event did or did not occur; for example, a team that wins a faceoff near their opponent's goal is more likely to score in the following seconds than they are to be scored upon, and have a higher probability of scoring than if their opponent had won the faceoff instead. For a player in a game, the sum of the weights of events in which they are involved can then inform us about that player's overall contribution to the game.

%While the use of the plus/minus count statistic is decades old, its use was originally restricted to a small number of teams. The popular account is that in its rawest form, Montreal 1950s, Toronto 1970s, league somewhere in that period.  %more on p/m here: Montreal did it on their own, Roger Neilson did it with the Leafs, the league eventually tracked it.

\subsection{Regression-Adjusted Measures}

The other most notable weakness of the standard Plus/Minus measure, or any of its derivatives, is coincident play: if two or more players are on the ice together for much of their shared time, it can be difficult to distinguish the abilities of each player from each other when so many of the outcomes to which they contribute are common to both. This problem is common to all goal-based team sports.

To handle this issue in basketball, \citet{rosenbaum2004mhnphttw} proposed to divide a National Basketball Association (NBA) game into intervals marked by the substitution of players onto the court. From this, he derived a number of independent events, each containing a number of scoring opportunities for each team. The outcome of each event is the difference in points scored between the two teams divided by the time elapsed during the interval; the predictors are indicators of the players on the court for each team -- positive for the home team, negative for the away team. Using a linear regression model of these player-predictors on the scoring outcome, each player's associated coefficient represents their contribution to the change in score in favor of their team; this is their ``adjusted plus/minus'' rating. Ideally, this measure will isolate a player's contribution to their own rating and remove it from others, as the quality of their teammates and their opponents is accounted for. %Do we need to add equations explicitly?

\citet{ilardi2008aprnaif2} modify this approach by taking every interval as not one but two events -- home scoring and away scoring -- and treating them as independent, conditional on the length of the event. Each player on the court appears in each of these two events, as an offensive and defense player respectively, and therefore has a distinct rating for each of these ``skills''; the combination of the two can then be taken as the total adjusted player rating.

Each of these procedures was conducted by \citet{macdonald2011rapsfnp} on NHL data by noting player substitutions from official game logs and using these to construct a table of events. \citet{gramacy2013epchrlr} considers a logistic regression model that focuses only on those events where one team scores a goal, which has the benefit of considering a much smaller set of events. Neither of these models allows for a user to simulate an entire game; the outcomes do not correspond to goal scoring processes, but to scoring rates in the former case and relative ability in the latter.

\subsection{Regularization Methods and Variable Selection}

One consequence of a regression modeling approach is the relatively large number of predictors against the number of events we can observe; in one season, there are roughly 400 different players in the NBA, and 1000 different players in the NHL. Because of this, estimates of ability on all players can be imprecise due to a potentially small sample on a subset of these individuals, through large variance or collinearity. One way to adjust for this is to regularize the estimates of each coefficient, producing biased estimates with lower variance. Ridge regression \citep{hoerl1970rrbefnp} is used by \citet{sill2010ina+uraot} for the NBA, and \citet{macdonald2012apfnpurr} for the NHL, to account for these difficulties; the degree of regularization was chosen through cross-validation on withheld observations. This approach, plus other Gaussian-derived models such as James-Stein estimation \citep{james1961eql}, are compared for the case of batting averages in \citet{brown2008ipbaftebabm}; this type of comparison is equally valid in this case.

%Ridge regression is equivalent to specifying a Bayesian model on the ensemble of parameters. In particular, for the linear model, each coefficient has its own independent Gaussian prior $\beta_i \sim N(0, \sigma_{\beta}^2)$, equivalent to a constraint on the squared sum of the coefficients, where the prior parameter $\sigma_{\beta}^2$ is either specified beforehand or selected using out-of-sample validation. The ridge method is in fact the simplest form of hierarchical model for this data; there is potential for more flexibility by assigning a hyperprior distribution to this parameter, such as the conjugate form $\sigma_{\beta}^2 \sim InvGamma(a,b)$, and obtaining a posterior estimate of each of the coefficients and the variance parameter simultaneously. 

Gaussian regularization methods produce estimates that are non-zero, but if the point is to distinguish the relative ability of two or more players, it may be that we are far less interested in the comparison between players ranked 499 and 500 than we would be between players ranked 1 and 2. Many of these lesser players may simply be nuisances for estimating parameters of greater interest. As a result, incorporating variable selection along with regularization may be useful. A standard method for this would be the Lasso \citep{tibshirani1996rsasvl}, in which we obtain both a subset of non-zero parameters as well as estimates for these parameters. %In this case, each coefficient has an independent Laplace prior, $\beta_i \sim Laplace (\lambda)$.

\subsection{Process Models}

The nature of substitution and scoring data from the NBA is vastly different from that of the NHL.  In the NBA, there are typically several scoring events for either team per rotation (the equivalent of a ``shift'' in hockey), and there are relatively few substitutions per game.  In the NHL, scoring events are much rarer, on the order of 10 minutes between goals, while players typically only spend about 30-60 seconds on the ice before returning to the bench for a substitution. As we show in Section \ref{s:goal-data}, roughly 98\% of these intervals have a total of zero goals scored. Using this linear regression approach, the event durations will not factor in, and significant information will be lost. Additionally, since the data are clearly non-Gaussian, methods based on Gaussian convergence properties may not be reliable, as the error terms and the prediction terms must be highly dependent to produce the majority-zero data.

The rarity of scoring events relative to the number of observable intervals suggests the use of a Poisson-type process model. Each event represents an observation of the same players on the ice, and any event that does not end in a goal is essentially censored by the change in players. This directly incorporates the observed duration of the event as well as accounting for the relatively sparse number of goals. Simple Poisson models have been used for making strategic decisions in hockey \citep{morrison1976otpgpmacsuih, beaudoin2010sfpgh}; these methods can be improved to account for heterogeneity in the scoring rate over time \citep{thomas2007itgih}.

Moreover, the game can often be divided into a number of discrete states that give additional information about the game. \citet{hirotsu2002umpmafmdotsatd} examine soccer as a continuous-time Markov process with 6 states: 2 teams can possess the ball on either half of the field, plus the state of having a goal scored in either net. \citet{thomas2006ippalihs} considers a larger state space for hockey with a semi-Markov process instead. Only when a team has possession of the ball/puck in their opponent's territory can they score a goal, so that this underlying state will then directly influence the scoring rate for each team. This method can be applied if data on location and possession is available, but this is not currently available to the public.

We expect that players in the game will similarly affect the scoring rates for each team. The Cox process model \citep{cox1972rmald} decomposes the rate of this process, described by the hazard function $h(t,X) = \lambda(t,X)$, into a time-varying component $\lambda_0(t)$ and a time-independent term for the inclusion of covariates $\lambda_x(X)$. Just as in the linear model case, these models can also be regularized, such as with the Lasso \citep{tibshirani1997lmfvscm}. %tweak this.

\subsection{Source of Data\label{s:goal-data}}

%In this research, we create a method for calculating statistically valid offensive and defensive player ability ratings for NHL forwards, defensemen, and goaltenders that take into account each of the principles listed above.  

Records of many National Hockey League (NHL) games are available to varying levels of detail. For the sake of dividing the game into discrete intervals, we use the interpretation of \citet{rosenbaum2004mhnphttw} and \citet{macdonald2011rapsfnp} that an interval should end either when a player substitution is made by either team or when an event occurs (e.g. when a goal is scored). This level of detail is available with ease in game records from the 2007-2008 season until the 2011-2012 season. We select those shifts in which both teams are at full strength -- each team has five skaters and one goaltender on the ice -- and note the duration of the event in seconds. The outcome is one of three possibilities: the home team scores, the away team scores, or neither team scores and at least one player substitution occurs. As Table \ref{t:event-counts} shows, over 98\% of the observations are non-goal outcomes, which is highly disproportionate compared to basketball.

For this analysis, we consider a process whose only events are goals scored by each team. We have additional information on shots on goal that did not result in goals, on penalties called that result in man-advantage situations, and on time-outs called (extremely rarely) by coaches. We do not include these at this stage to keep the analysis on events that directly influence the final result of winning or losing the game, since shots on goal only lead to goals a fraction of the time, and the relationship between shots on goal and goals is not as simple as a fixed fraction of events. Any processes that lead to shots must also lead to goals, and to add additional competing processes to the model would add an additional level of complexity that is beyond the scope of this investigation. (See \citet{macdonald2012egmfentap} for how shots can be used in a standard regression scheme.)

\begin{table}
\begin{center}
\caption{A count of the events of each type in the database. A home team advantage is apparent. \label{t:event-counts}}
\begin{tabular}{c|ccc}
\hline
\hline
Seasons: 2007-2012 & Away Goal & No Goal (Changes) & Home Goal \\
\hline
Total Events & 10,935 & 1,301,799 & 11,981 \\
Percent of Total Events & 0.83 & 98.27 & 0.90 \\
\hline
\hline
\end{tabular}
\end{center}
\end{table}

For each season, we divide the games into two groups, uniformly at random -- one for in-sample training (all observations from 80\% of the games) and one for out-of-sample validation (the remaining 20\% of games). When we perform any tuning parameter selection, we further subdivide the in-sample training set for cross-validation.

% -- for the purposes of cross-validating the regularization parameters.

%Number of distinct players per season, in total; players appearing in all 4 seasons, etc.

%Our outcome variable is difficult to model using typical classification approaches:
%-histogram of interval time, goals and no-goals.
%(**Shift Length EDA**)
%(**Home vs. Away EDA**)
%(**Score Differential EDA**)

%%%%%%%%%%%%%%%%%%%%%%%%%%%%%%%%%%%%%%%%%%%%%%%%%%%%%

\section{Model Specification}\label{s:model-spec}

We model the stochastic nature of the game as a model of two competing processes for the scoring of a goal, censored by player substitutions. Each process has parameters for offensive and defensive characteristics, and these parameters are regularized by partial pooling. We use penalized maximum likelihood and full hierarchical Bayesian models to infer parameters of interest.

\subsection{Events Obey A Competing Processes Model}

There are, at a minimum, two opposing processes in a hockey game: the home team tries to score on the away team, and vice versa. Both of these events are relatively rare compared to the number of observed event intervals, so that it is natural to model these as competing stochastic processes. Predictors that modulate these processes can be the teams in the game, the score of the game, the players on the ice, or some other combination, and the same predictors appear in each process. 

We choose a Cox proportional hazards model for each process, so that the hazard function has separate components for time dependence and predictors, as $h(X,t) = h_0(t)h_1(X)$, where $X$ can represent various factors such as the players and/or team on the ice. For this investigation we begin with $h_0(t)=1$; more information on the location of the puck at each $t=0$ may allow us to refine the time-based component in future investigations.

From this, each team's scoring rate is modeled as a log-linear Poisson process. The intercept terms, labeled $r^h$ and $r^a$, represent the baseline scoring rates for the home and away teams, since as we see in Table \ref{t:event-counts}, the overall scoring rate for the home team is greater than for the away team; in this way, we explicitly detect a home-ice advantage. For each predictor indexed by $p$, let $(\omega_p, \delta_p)$ be a measure of the offensive and defensive contribution for that predictor, so that a rating of zero corresponds to an ``average'' contribution; the corresponding indicators are $X_p^h$ and $X_p^a$. 

The scoring rates for each process are 

\begin{eqnarray*}
\lambda^h & = & \exp(r^h + \sum_p (X_p^h \omega_p + X_p^a \delta_p)); \\
\lambda^a & = & \exp(r^a + \sum_p (X_p^a \omega_p + X_p^h \delta_p)) 
\end{eqnarray*}

for this combination. For each instance of this process, $T^h$ and $T^a$ are the times to each event for these processes, and let $t$ be the first time at which any players on the ice are substituted, thereby censoring the scoring process. We assume that the (unmodeled) censoring time is independent of these event times, and that conditional on the predictors, these events are independent of each other. The outcome can then be registered as 

\[ Y = \left\{ 
\begin{array}{cc} 1 & if\ T^h\ <\ T^a,\ T^h\ <\ t \\ 
-1 & if\ T^a\ <\ T^h,\ T^a\ <\ t \\ 
0 & otherwise \end{array}    
\right. 
\]

\noindent so that $(1, 0, -1)$ represents a home goal, no goal and away goal respectively. Let $T = \min\{t, T_h, T_a \}$ be the observed time of the event.

Because of the independence condition, the likelihood for this event is then the product of the individual likelihoods, noting if either or each of the events was censored. With the survival function form $S(x) = P(T>x)$, we have

\begin{eqnarray*}
f(Y|\lambda_h, \lambda_a, T) & = & f_h(T|\lambda_h)^{\mathbb{I}(Y=1)}S_h(T|\lambda_h)^{\mathbb{I}(Y \neq 1)} \ \times \\
& & f_a(T|\lambda_a)^{\mathbb{I}(Y=-1)}S_a(T|\lambda_a)^{\mathbb{I}(Y \neq -1)}. 
\end{eqnarray*}

Using this approach, each predictor's offensive parameter coefficient represents the change in the team goal scoring rate with respect to a baseline rate (in particular, if they are replaced by another player of typical ability), and likewise for their defensive parameter and the opposing goal rate.

This method has several advantages for this class of data. Rather than trying to model a single outcome, such as goal differential, we can simultaneously calculate both the offensive and the defensive player ability parameters for each player, which are known to be distinct. The parameters we calculate have a meaningful interpretation in terms of game outcomes, since it reflects an increase or decrease in scoring rate. We can assess a player's marginal goal fraction over data in question by comparing the expected number of goals scored and allowed by their team given their ratings against the same data with ratings set to zero.

In addition to the offensive and defensive abilities of each player, we can account for several other possible influences. We can fit parameters to a whole team to capture their average ability, rather than simply including all the players independently. If we include both teams and players as predictors, this would change the interpretation of a ``player effect'' to be relative to the performance of one's team. We can also model an effect for the in-game score differential, since many teams may change their offensive and defensive strategies depending on how far ahead or behind they are in the game. This may best be accomplished by selecting a different intercept term depending on the score.

\subsection{Regularization of Parameter Estimates}

Even though we observe hundreds of thousands of discrete shift intervals in a season, the potential number of parameters in this model is also very large, and many of the player ability measures will be made with only a small number of observations, such as players who appear in only one game. Worse yet are those players who are not on the ice for any goal by one team and therefore have a maximum likelihood estimate of minus infinity for each of their parameters. To account for this, we use a hierarchical model to shrink parameter estimates toward a common mean (namely, zero), with the possibility that different positions (center, goaltender, winger and defenseman) have different shrinkage behavior. We have a number of choices for how to carry out this {\em regularization}: the choice of prior distribution or penalty term, the degree of hierarchical structure we impose, and whether we choose to minimize a function or integrate over a distribution.

The two standard choices for a prior/penalty distribution are the Gaussian and the Laplace, which penalize the mean squared error and absolute error respectively. We can also consider a third class that joins the two, in the spirit of the Elastic Net method \citep{zou2005regularization}, the Laplace-Gaussian distribution:

\vskip 0.5cm

\begin{center}
\small
\begin{tabular}{cc}
\hline
Prior Type & PDF \\
\hline
Lasso/L1 &  $f(x|\lambda) = \frac{\lambda}{2} \exp (-\lambda |x|)$ \\  %Laplace$(\lambda)$ &
Ridge/L2 & $f(x|\sigma^2) = \exp (-x^2/(2\sigma^2)) / \sqrt{2\pi\sigma^2}$ \\  %& Gaussian$(0, \sigma^2)$ 
Elastic Net/L1+L2 & $f(x|\lambda, \sigma^2) = \frac{\exp(-\sigma^2 \lambda^2/2 -\lambda |x| -x^2/(2*\sigma^2))}{\sqrt{8 \pi \sigma^2} \Phi(-\sigma \lambda)}$ \\   %& Laplace-Gauss$(\lambda, \sigma^2)$ 
\hline
\end{tabular}
\end{center} %\footnote{$Phi(x)$ is defined as the standard Gaussian CDF.}

\vskip 0.5cm

While each of these regularization options act to stabilize parameter estimates, both in cases with few observations and in those pairs or multiples with high collinearity, each family gives a different interpretation for the shrinkage behavior of the covariates.

If we choose the L1 method and set each $\lambda$ to a constant, then we have a (relatively standard) Lasso implementation, in which the penalized MLE or MAP estimates for the parameter may be exactly zero with non-zero probability, which yields a smaller subset of predictors for which the scoring rate change is distinguishable from zero. The L2 method with constant $\sigma^2$ terms yields a ridge regression-like result, in which the penalized MLE or MAP estimates for each parameter are brought closer but not exactly to zero. Compromising with the L1+L2 method allows for some of the benefits of both properties, but may sacrifice the ease of implementation that can be found in the simpler cases. In the case of simple optimization, the L1 and L2 cases are suited to using cross-validation to choose the penalty weights $\lambda$ and $\sigma^2$. If we are considering multiple partially pooled groups, cross-validation may no longer be computationally feasible, since searching the space of possible parameters becomes more difficult the more dimensions we add. 

%Neither of the Gaussian or Laplace distributions is conjugate to the parameters in the likelihood, making direct sampling of the full conditional distributions more computationally intensive. For the L1 and L2 cases, the scale parameters have semi-conjugate priors, meaning that we can sample each of these terms from their full conditionals directly. This advantage disappears with our Elastic Net prior, though the vastly smaller number of terms to consider, compared to the full conditional distributions of $\omega$ and $\delta$, allows for direct draw approximations rather than Metropolis-type sampler steps.

\subsection{Implementation}

We have two types of problems that we consider: those in which the total distribution of predictors and their group-level variance terms is of direct interest, and those in which we are only interested in selecting a subset of relevant predictors. The former case requires simultaneous estimation of a number of shrinkage parameters, and this dimensionality makes a search of the space difficult to accomplish with cross-validated methods, so we use the full hierarchical Bayesian approach. In the latter case, there is typically only one dimension of interest, as we wish to select from only one relevant subset of predictors, and so here we can use penalized maximum likelihood estimation much more easily.

\subsubsection{Optimization of Penalized Likelihood}

We use maximization of a penalized likelihood to get rough parameter estimates, with modest levels of L1 and/or L2 shrinkage to handle parameters with minimal information in the data, such as players who played in only one game. We can use this as a starting point for Markov Chain Monte Carlo to obtain estimates for the pooled variance/shrinkage parameters. For each MCMC routine, we discard a sufficient number of initial samples as burn-in and thin the chain sufficiently so that the thinned chain has negligible autocorrelation for all parameters and a sufficient number of uncorrelated samples (in each of our cases, a minimum of 500) for use in inference.

We can also simply scan through a series of values for each shrinkage parameter, selecting the optimal value through out-of-sample validation. This is easiest when there is only one shrinkage parameter to estimate.

\subsubsection{Full Posterior Estimation with MCMC}

The full hierarchical model has three levels: from the data, to the predictor coefficients, and finally to their partial pooling prior distributions. We use a Gibbs sampler blocked on pairs of variables to estimate model parameters.

\begin{itemize}

\item \textbf{Level 1}: Each outcome $(Y|X^h,X^a,\omega,\delta,t)_i$ is distributed as the competing process model. Each predictor block $(X_i^h, X_i^a)$ is stored as a sparse vector, given that there are typically no more than 16 total non-zero terms in each row.
\\

\item \textbf{Level 2}: Each coefficient pair $(\omega, \delta)_p$ is distributed according to its prior distribution. In the Laplace-Gaussian case, this has four terms corresponding to the group $g(p)$ that has predictor $p$ as a member: the Laplace terms $(\lambda_{\omega, g}, \lambda_{\delta, g})$ and the Gaussian terms $(\sigma^2_{\omega, g}, \sigma^2_{\delta, g})$. 

As the intercept terms $r^h$ and $r^a$ effectively correspond to their own $(\omega, \delta)$ pair and belong to their own group, each acts as their own group mean; weak hyperpriors on their own prior terms act marginally as weak prior distributions.

Each pair $(\omega_p, \delta_p)$ is updated using a Metropolis sampler with a bivariate Gaussian proposal distribution. Indexing each observed shift with $i$, the target distribution 

\[ f(\omega_p, \delta_p | Y,X, \sigma_{\omega, g(p)}, \sigma_{\delta, g(p)}, \lambda_{\omega, g(p)}, \lambda_{\delta, g(p)}) \] 
equals the product 

\begin{eqnarray*} 
f(\omega_p, \delta_p | \sigma_{\omega, g(p)}, \sigma_{\delta, g(p)}, \lambda_{\omega, g(p)}, \lambda_{\delta, g(p)}) 
\prod_{i: p \in (X_i^h, X_i^a)} f(Y|X^h,X^a,\omega,\delta,t)_i. 
\end{eqnarray*}

We initialize all $(\omega,\delta)$ terms with a penalized maximum likelihood estimate using relatively loose shrinkage parameters.
\\

\item \textbf{Level 3}: Each Laplace $\lambda$ term has a weak Gamma conjugate prior; each Gaussian $\sigma^2$ term has a weak Inverse Gamma conjugate prior. If the Laplace-Gaussian is used, these priors are no longer conjugate to their respective parameter forms.

Each pair $(\lambda_{\omega,g}, \sigma^2_{\omega,g})$ is updated through a pair of univariate grid approximation samplers. The first samples according to the density along the sum of approximate total shrinkage, $1/\sigma_{\omega,g} + \lambda_{\omega,g}/\sqrt{2}$, while keeping the relative fraction of shrinkage $\frac{\lambda_{\omega,g}/\sqrt{2}}{\lambda_{\omega,g}/\sqrt{2} + 1/\sigma_{\omega,g}}$ constant;\footnote{The $\sqrt{2}$ factor is added to reflect the fact that a Laplace distribution with scale 1 has a variance of 2.} after updating these values, the second samples the relative fraction while keeping the approximate total constant. This is repeated for each pair $(\lambda_{\delta,g}, \sigma^2_{\delta,g})$. (One can always sample directly from the bivariate grid approximation as well, though this is less computationally efficient.)

\end{itemize}

We constructed the sampler using the \texttt{R} programming language with supporting back-end code in \texttt{C++}. Execution time varies with the total number of covariates, with the simplest cases (200,000 outcomes and 60 covariates) taking 30 processor-minutes, to the more complicated runs (200,000 outcomes and 2600 covariates) requiring roughly 60 processor-hours. We used multiple parallel chains with sufficient burn-in periods to collect a sufficient number of uncorrelated samples. We validated the sampler using the method of posterior quantiles \citep{cook2006vsb}.

In each of these cases, we can judge the performance of each selected model initially using in-sample measures, then confirming goodness of fit by checking against our held-out data. For MCMC, we use the Deviance Information Criterion, calculated using the individual samples and the average over all samples, applied to the likelihood of the original (fitted) data for in-sample fit, as well as to our withheld data for out-of-sample validation.

\section{Analysis of Full Posterior Distribution}\label{s:applications}

%This modeling framework allows us to investigate many different questions about the game and the distribution of playter abilities. We make use of either penalized maximum likelihood estimation or full hierarchical Bayesian inference depending on the nature of the quantity under investigation; we tend to use the latter when the group-level variability is a direct quantity of interest rather than simply a tool for shrinkage. 

Since all analyses in this investigation are conducted on events where both teams are at full strength, we refer to any particular coefficient pair $(\omega_p, \delta_p)$ as the Mean Even Strength Hazard (MESH) rating for the corresponding predictor, such as the team (as in Section \ref{s:team-rating}), a particular player (Section \ref{s:grand-model}), or the extra contribution of a pair of players (Section \ref{s:player-pairs}). We estimate the net MESH rating as offensive ability minus defensive liability, $\omega_p - \delta_p$.

\subsection{Home-Ice Advantage and Game Score}\label{s:homeice}

The simplest version of this process model has only two coefficients, the intercepts for the home team and away team processes:

\[ \lambda^h = \exp(r^h); \qquad \lambda^a = \exp(r^a). \]

\begin{figure}  
\begin{center}
\includegraphics[width=\linewidth]{\imloc 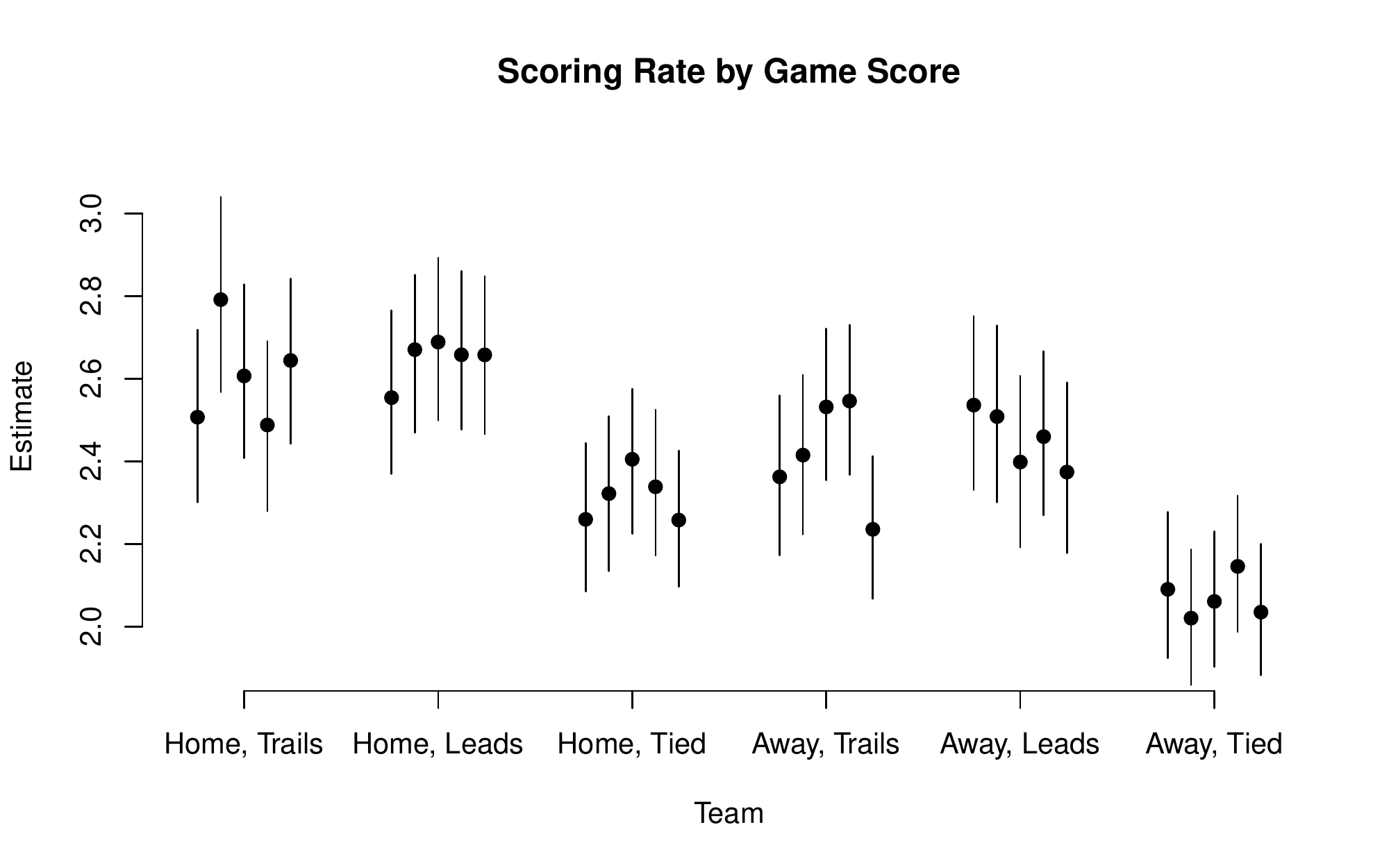}  %, natwidth=1000, natheight=500
\end{center}
\caption{The scoring rates per 60 minutes for generic home and away teams in each individual season, divided by game score. Points are posterior means, lines are central 95\% credible intervals. The home team consistently outscores the away team in all five seasons and overall within each game situation; scoring rates for both sides are elevated when the game is not tied, whichever team is winning. \label{f:intercepts}}
\end{figure}

We can extend this by specifying different intercepts for different game score situations. For this analysis we choose three: when the home team is winning, tied or trailing.\footnote{While we can extend this more generally to all combinations of game score, the results of this division are quite robust.} Figure \ref{f:intercepts} shows the estimates of these intercepts in each of the five seasons under consideration, for home and away teams, by taking each $\exp(r^h)$ and $\exp(r^a)$, the per-second rates, and multiplying up to a full (hypothetical) 60-minute game.

It is clear that the home team has a consistent advantage. Whether or not the effective home scoring rate is actually identical in each of the five seasons, they are so close as to be indistinguishable from each other; this is similar for the away scoring rate. The year-to-year variability in home and away mean rates is consistent with a common goal-scoring rate across all five seasons; simulations verify that the change in estimated means is consistent with the spread in estimation based on the generation of a season's worth (1230 games) of goals for each team from the Poisson model.

It is also clear that there is a change in scoring rates by game score. Interestingly, the scoring rates for each team are raised by the same amount when a team is leading or trailing, compared to when the score is tied. This suggests that teams are more cautious during tie scores, and that efforts by the trailing team to increase their own scoring rate, or by the leading team to increase their margin, result in a corresponding and roughly equal increase in their opponents' rate.

\subsection{Overall Team Performance, Per Season}\label{s:team-rating}

Because each of the 30 teams in the data is present in roughly one fifteenth of the total events, we do not expect the degree of sparsity as when we model the impact of individual players. This does not mean, however, that the model cannot benefit from partial pooling on team parameters, both to reduce the effective dimensionality of the model and to improve predictive accuracy. This model is then specified as

\[ \lambda^h = \exp(r^h + \omega_{home} + \delta_{away}); \qquad \lambda^a = \exp(r^a+ \omega_{away} + \delta_{home}) \]
with partial pooling under one of our chosen schemes; in general, this is of the form

\[ \omega_{team} \sim \mathrm{Laplace-Gaussian}(\lambda_{team}, \sigma^2_{team}) \]
where the shrinkage behavior depends on the prior specification for the parameters $(\lambda_{team},\sigma^2_{team})$. We include three sets of intercepts for game score situation, though additional analysis shows that our parameter estimates are insensitive to this model choice.

\begin{figure}
\begin{center}
\includegraphics[width=\linewidth, natwidth=1000, natheight=500]{\imloc 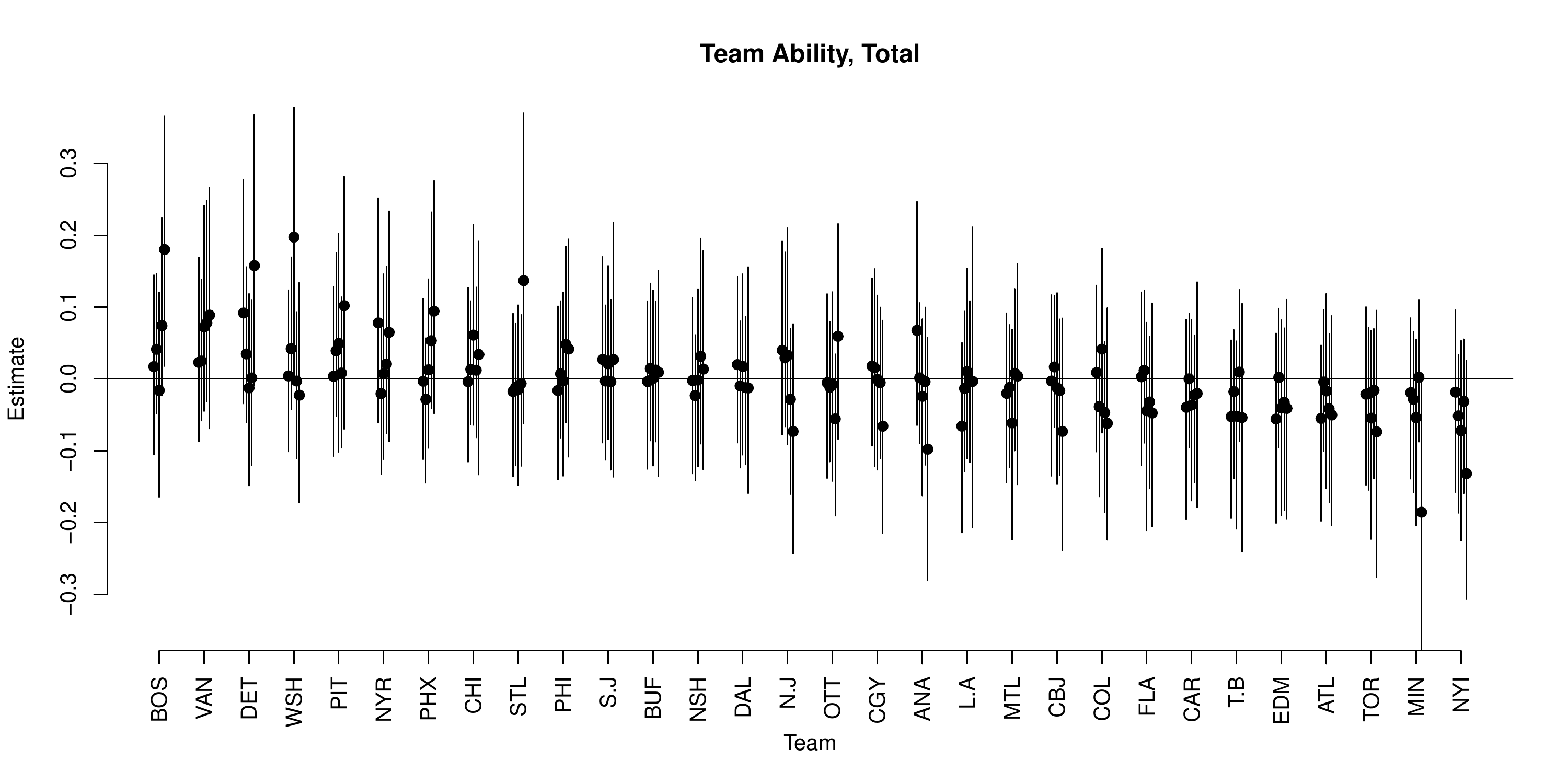}
\end{center}
\caption{Total team ability estimates for each team in the NHL, grouped by team for each season; order is by overall team rating. Points are posterior means, lines are central 95\% credible intervals. A rating of 0.1 corresponds to a differential of roughly 0.3 goals per game scored or prevented. Note that only two team-years, the 2012 Boston Bruins and 2010 Washington Capitals, have effects that are significantly different than average. \label{f:teams-total}}
\end{figure}

\begin{table}
\begin{center}
\caption{DIC for the Laplace, Gaussian and Laplace-Gaussian pooling priors for the model with teams as explanatory variables. The Laplace-Gaussian performs the best in each season and overall in both the in-sample and out-of-sample cases.\label{t:team-DIC}}
\begin{tabular}{ccccccc}
\hline
\hline
& & Insample & & & Outsample & \\
Season & L1 & L2 & L1+L2 & L1 & L2 & L1+L2 \\
\hline
2007-2008 & 57701 & 57703 & \textbf{57691} & 14088 & 14088 & \textbf{14085} \\
2008-2009 & 59996 & 59966 & \textbf{59962} & 15071 & 15064 & \textbf{15064} \\
2009-2010 & 61415 & 61359 & \textbf{61348} & 15710 & 15704 & \textbf{15702} \\
2010-2011 & 62552 & 62521 & \textbf{62515} & 15541 & 15538 & \textbf{15537} \\
2011-2012 & 62398 & 62398 & \textbf{62377} & 15983 & 15983 & \textbf{15982} \\
\hline
\hline
\end{tabular}
\end{center}
\end{table}

We estimate these parameters within each season using MCMC for each of the three submodels for pooling. For each shrinkage mode, two variance components are estimated, for total offensive and defensive ability respectively. For the Laplace-Gaussian prior form, there are four total parameters, rather than two, and this mode has the lowest Deviance Information Criterion for all five seasons, both in- and out-sample, as shown in Table \ref{t:team-DIC}. From this point on, we focus on results using only the full Laplace-Gaussian prior.

Figure \ref{f:teams-total} shows the posterior distributions for each team's net MESH rating within each season, using the Laplace-Gaussian prior. As expected, these track well with the number of goals scored and allowed by each team during these seasons, since the correlation of parameters across teams is minimal; teams play each other no more than eight times per season out of a total of 82 games. There are also several significant deviations for some teams for one season compared to the rest, such as St. Louis in 2012 (very positive) and Minnesota in 2012 (very negative), that are not statistically distinguishable from their other performances but still illuminating nonetheless.

It is worth noting that very few of these parameters have 95\% credible intervals that do not contain zero, suggesting that the amount of information to distinguish a team from being truly ``average'' is quite small; however, it is clear that some teams are highly probable to be better (or worse) than other teams in the league, so that distinguishing a statistically significant ordering is within the reach of this model. Given the limited amount of information, the question arises as to whether we can reliably distinguish player abilities from average when the information on a single player is much less than that for a single team. We address this question in the following section.

\subsection{Distribution of Player Abilities, Across All Seasons}\label{s:grand-model}

\begin{figure}[!ht]
\begin{center}
\includegraphics[width=\linewidth, natwidth=1000, natheight=500]{\imloc 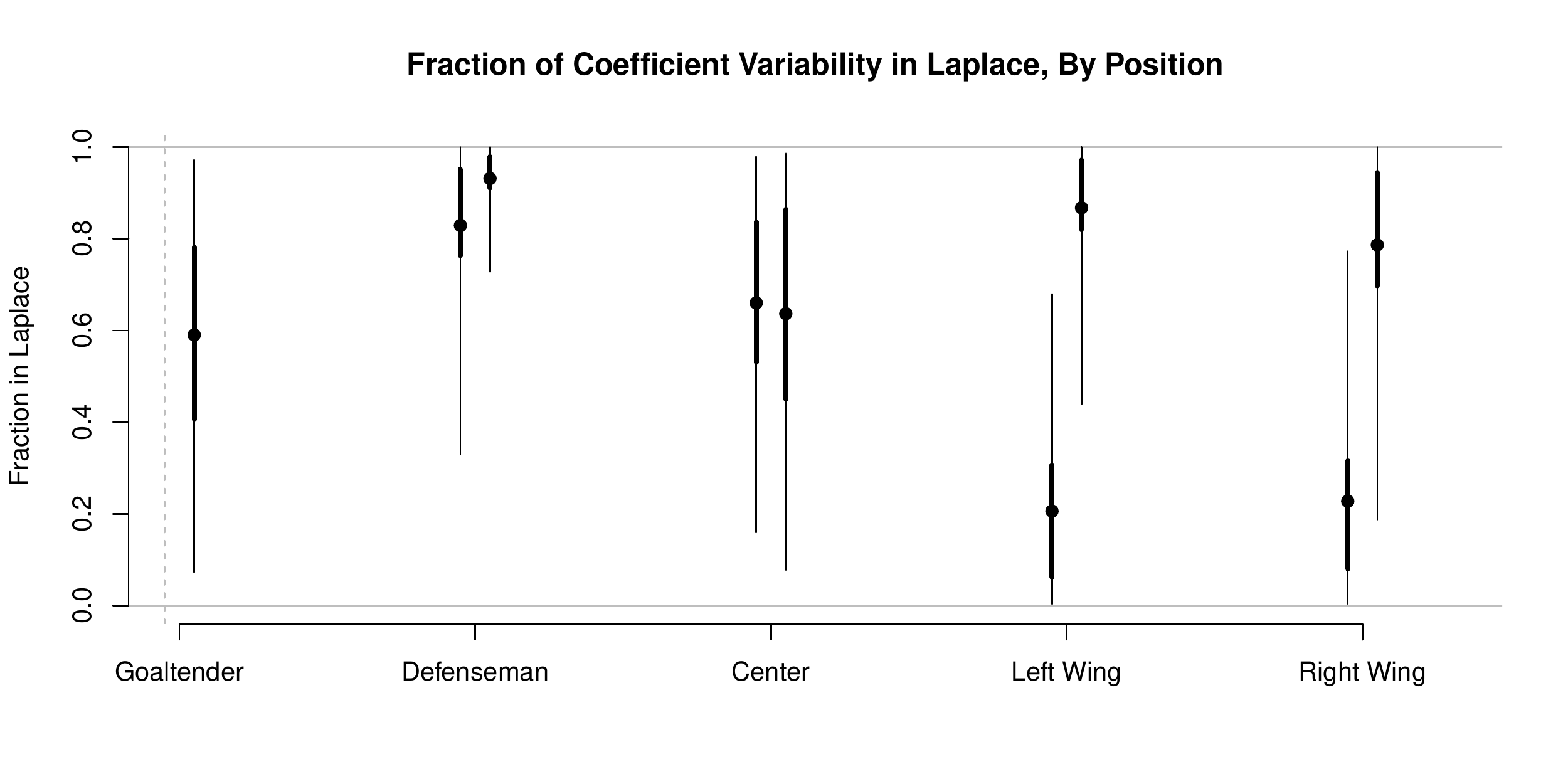}
\includegraphics[width=\linewidth, natwidth=1000, natheight=500]{\imloc 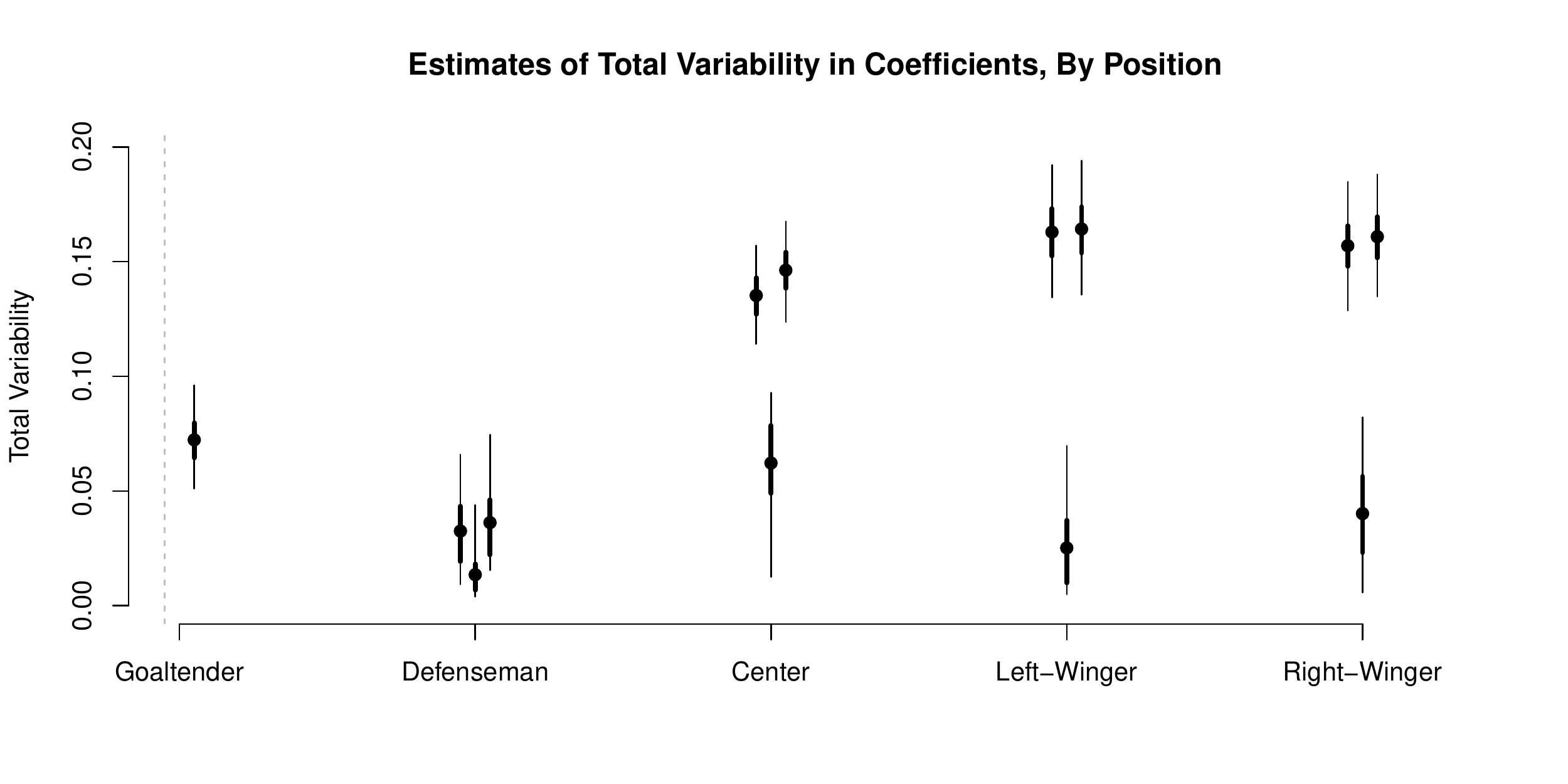}
\end{center}
\caption{Variability properties of coefficient estimates by position. Thick and thin lines represent 50\% and 95\% credible intervals; points represent estimated means. \textbf{Top}, the approximate fraction of the variability that can be attributed to the Laplace component of the Laplace-Gaussian error distribution, for each position and for offense and defense. For most positions there is a strong tendency towards the Laplacian distribution, with heavier tails and more outliers; this is less pronounced for winger offense. \textbf{Bottom}, the offensive, defensive and total variability by position. Goaltenders have only defensive variability, which is considerably more variable than defense for any skating position. Offensive players (centers and wingers) have more variability in offense, and every skating position has minimal variability in defence. \label{f:variabilities}}
\end{figure}

The estimation procedure for team effects is relatively straightforward, given the relative balance of the design matrix. Once we consider individual players, more questions arise since the design matrix can be far more unbalanced; for example, a player's defensive rating may be trickier to estimate because they share the majority of their shifts with a single goaltender. Arguably, it gets worse if both players are {\em great} players, since they may both be retained by a single team for much of their careers.

This is made easier when dealing with data from multiple seasons, as the more players change teams, the more the players in the league will mix. We therefore model player abilities as constant over all five seasons, which we refer to as the ``grand model'', specified with the following terms:

\begin{itemize}

\item Overall home and away effects with score differential effects.

\item Offensive and defensive parameters for all skaters (centers, wingers and defensemen).

\item Defensive parameters only for goaltenders.

\item Laplace-Gaussian pooling for each type of ability and each position parameter (center, left wing, right wing, defenseman, goaltender).

\end{itemize} 

We do not include team effects at this stage specifically because we are trying to compare players across teams, and their collinearity with goaltenders is needlessly complicating. We are still resigned to the degree of confounding in defensive estimates, since the goaltender not only plays a large role, but is not typically replaced throughout the game, most often only relieved during a poor outing. We use the standard MCMC implementation to estimate parameters.

There is only a small subset of players whose ratings can be considered statistically significant. Of 1592 total players over five seasons, 37 have player ratings whose effective total (offensive skill minus defensive liability) have 95\% central credible intervals that do not contain zero. Of these, 36 are positive; only one player, Stephane Veilleux, had a negative total rating with statistical significance, suggesting that he is a good enough player to log regular ice time with a major league team, but not so good that his contributions in even strength are less than the league average. (This is not necessarily the same as a ``replacement''-level player.) The top five players at each position group are given in Table \ref{t:grand-players}.

\subsubsection{Overall Variability of Rating By Position}

%\begin{minipage}
\begin{figure}[!ht]
\begin{center}
\includegraphics[width=\linewidth]{\imloc 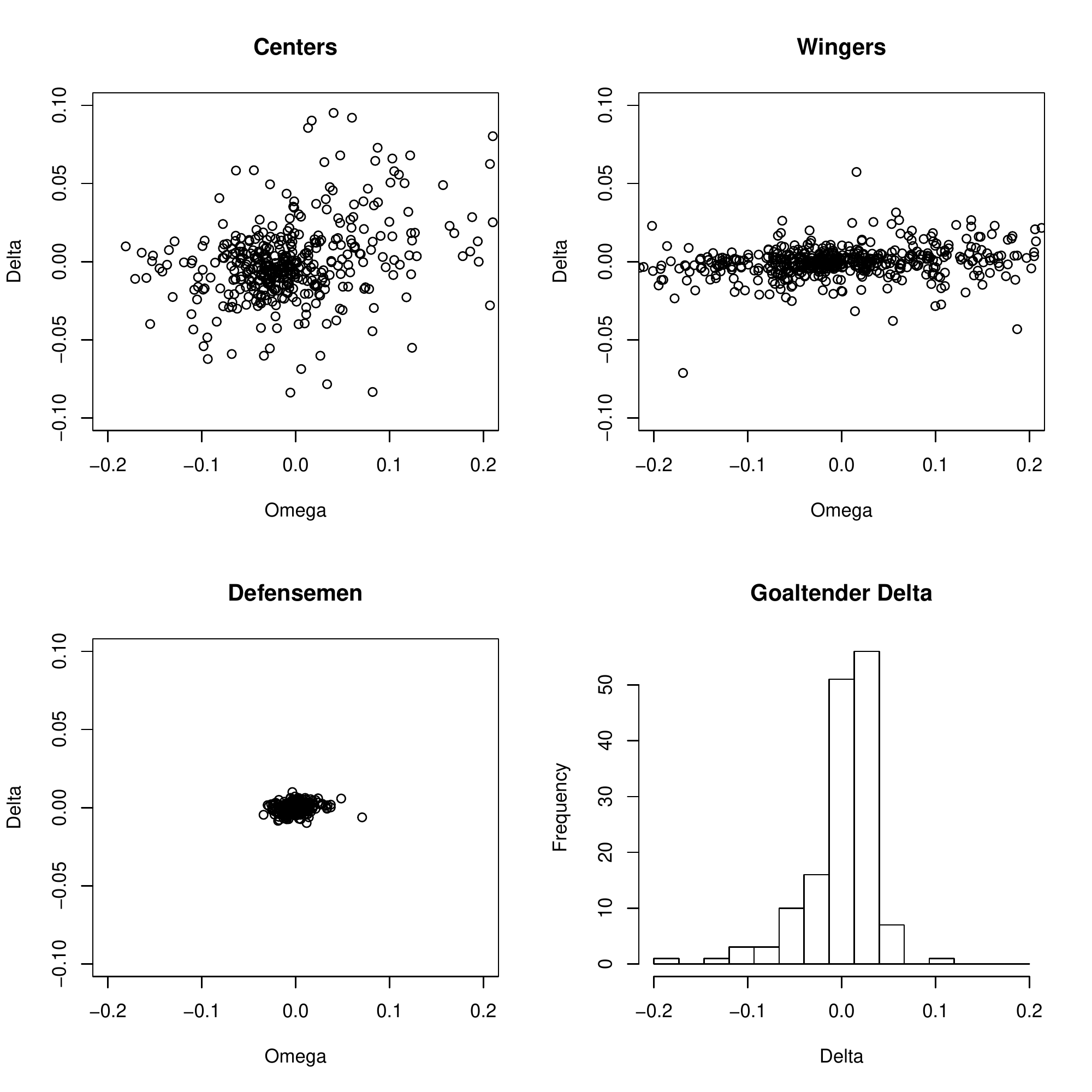}
\caption{Scatterplots of player ability estimates by position. There is little if any correlation between a player's estimates of offensive and defensive ability. \label{f:scatterplot-pos}}
\end{center}
\end{figure}

Figure \ref{f:variabilities} shows the variability of player abilities at each position according to their respective Laplace-Gaussian distributions. The first graph shows us an approximate proportion of the fraction of variability best explained by the Laplace term, as an indicator of the degree to which a distribution of players has heavier tails; the higher this is, the higher the number of ``extreme'' players. The second graph shows the total variability of player abilities as the standard deviation of player estimates at each iteration of the MCMC. 

Several matters are apparent. There is considerable variability in offensive ability for forwards (centers and wingers) but far less for defensemen. This is consistent with the notion that defensemen have less impact on offensive output during even-strength situations.

For all positions other than goaltender, defensive variability is far smaller than it is for offense. Two explanations are immediate. First, it may be that the collinearity between skaters and goaltenders is causing our estimates of goaltender ability to be more variable than they are in reality, and less variable for the skaters. Second, since the total defensive burden is shared by six players (five skaters plus one goaltender) rather than the five for offense, and the bulk of defensive skill is taken up by the goaltender, the total amount of ``defensive skill'' available to be shared by skaters is considerably smaller, and therefore there is less total variability between players.

How valuable is an individual position to a team? A typical starting goaltender plays about 60 full games a season for their team, while first-line offensive and defensive players will have the equivalent of roughly 30 and 35 full games respectively. On average, a good goaltender is worth roughly what a good offensive player is to a team's total output with respect to ``average'' players, while a good defensive player appears to be worth considerably less.

The center position has, on the whole, more effect on defensive performance than a defenseman does, and wingers seem to have roughly equal defensive variability as the defense position has total variability. This would seem to confirm the case that when forwards have control of the puck, particularly in their offensive zone, they deny the likelihood of their opponents being able to score. As we show soon, this does not mean that a player with a high $\omega$ rating must therefore have a high $\delta$ rating.

From these overall results, we move on to describe the individual performances of players over the five-season period, as organized by position. Table \ref{t:grand-players} lists the top five players in each position group under the grand model; we provide a more complete list of players at each position in the supporting material, including several of the worst players at each position.

\begin{table}[h]
\begin{center}
\caption{Top five players at each position, by overall rating, over five NHL seasons (2007-2012). Listed are mean ratings, 95\% credible intervals, and posterior probabilities that the player is the best at their position. \label{t:grand-players}}
\begin{tabular}{cccc}
\hline
\hline
Player & Total MESH  & 95\% Credible & \% Probability \\ 
&Rating & Interval & Best Player \\
\hline
& \textbf{Center} & & \\
Pavel Datsyuk & 0.463 & (0.262, 0.668) & 39.5 \\
Sidney Crosby & 0.388 & (0.155, 0.598) & 18.1 \\
Henrik Sedin & 0.355 & (0.096, 0.606) & 13.3 \\
Patrice Bergeron & 0.280 & (0.075, 0.535) & 8.7 \\
Evgeni Malkin & 0.266 & (0.048, 0.429) & 4.5 \\
\hline
& \textbf{Winger} & \\
Alexander Semin & 0.321 & (0.167, 0.459) & 3.9 \\
Alex Ovechkin & 0.318 & (0.160, 0.478) & 6.6 \\
Marian Gaborik & 0.308 & (0.128, 0.478) & 7.6 \\
Loui Eriksson & 0.258 & (0.097, 0.407) & 6.0 \\
Alexander Radulov & 0.249 & (0.003, 0.490) & 5.5 \\
\hline
& \textbf{Defense} & \\
Zdeno Chara & 0.077 & (-0.015, 0.244) & 11.7 \\
Mark Streit & 0.0427 & (-0.038, 0.207) & 6.1 \\
Jaroslav Spacek & 0.0373 & (-0.033, 0.163) & 4.5 \\
Mike Green & 0.036 & (-0.031, 0.185) & 2.6 \\
Matt Carle & 0.034 & (-0.026, 0.161) & 3.2 \\
\hline
& \textbf{Goaltender} & \\
Henrik Lundqvist & 0.186 & (0.076, 0.292) & 36.0  \\
Tim Thomas & 0.120 & (0.005, 0.233) & 20.6  \\
Jonathan Quick & 0.102 &  (-0.012, 0.221) & 14.2  \\
Martin Brodeur & 0.101 & (-0.009, 0.209) & 7.0  \\
Roberto Luongo & 0.100 & (-0.010, 0.211) & 5.3  \\
\hline
\hline
\end{tabular}
\end{center}
\end{table}
%\end{minipage}

\subsubsection{Assessing Model Fit}

We compare the models with basic intercepts, team parameters and player parameters by calculating the likelihood of points withheld from the original model fit, in each of the five seasons and altogether, with respect to the posterior mean for each parameter. As we show in Table \ref{t:model-fit-comparison}, the likelihood is highest in all five seasons for the player parameter model, even with a much higher number of parameters.

\begin{table}
\begin{center}
\caption{Comparing the out-of-sample doubled negative log likelihood for the models for game score only, team parameters and player parameters respectively. Even with a large number of extra parameters, the model with player effects yields considerably better fit for the withheld data than the alternatives. \label{t:model-fit-comparison}}
\begin{tabular}{cccc}
\hline
Group & Score & Team & Player \\
\hline
2007-2008 & 14096.1 & 14088.7 & \textbf{14002.5} \\
2008-2009 & 15077.2 & 15072.1 & \textbf{15012.1} \\
2009-2010 & 15704.7 & 15702.1 & \textbf{15643.1} \\
2010-2011 & 15543.7 & 15542.1 & \textbf{15488.5} \\
2011-2012 & 15994.8 & 15992.3 & \textbf{15940.1} \\
Total & 76416 & 76397 & \textbf{76087} \\
\hline
\end{tabular}
\end{center}
\end{table}

The adequacy of the fit of the model to data is harder to assess. The process is inherently noisy -- the number of goals in a game for any team varies wildly -- and so our ability to predict the behavior of any one game is minimally improved when adding player parameters. To check the adequacy of our estimates for player parameters, we simulate data for each game in the withheld set using the posterior mean and check the sum of the goals scored by the home and away teams in each simulated season against the truth; we find that the true data lies within the 95\% simulated confidence interval each time, and with every model (score alone, teams, and players respectively).

\subsubsection{Players That Make The Greatest Total Difference}

Since the ratings represent multipliers to the default scoring rate, we can quickly estimate the total contribution of a player over the observation period as the difference in expected goals, scored and allowed by any average team, relative to an average player,

\[ G_{net} = \left[\left(\exp(r_{base} + \omega_p) - \exp(r_{base})\right) - \left(\exp(r_{base} - \delta_p) - \exp(r_{base})\right)\right]\times T_{total,p}. \]

A mean intercept parameter $r_{base} = -7.3$ corresponds to roughly 2.4 goals per 60 minutes. Table \ref{t:grand-players-goals} lists the top 20 total goal producers and preventers over the five season period. Four goaltenders make the top 20 list; despite the fact that defensemen typically log more ice time than forwards, no defencemen make the top 20. We can adjust these ratings to reflect teammates and opponents by using the expected goals in each shift given all other player ratings, to handle nonlinearity in the rate relationship.

\begin{table}
\scriptsize
\begin{center}
\caption{The top 20 even-strength players in the NHL over 5 seasons (2007-2012) according to the net number of goals scored or prevented $G_{net}$, assuming a baseline scoring rate of roughly 2.4 goals per team per 60 minutes. At position 81, Zdeno Chara is the highest-ranked defenseman in this time period.\label{t:grand-players-goals}}
\begin{tabular}{cccccccr}
\hline
Rank & Player & Pos & Time (s) & +Scored & +Stopped & $G_{net}$ & \%Pr(Best)\\
\hline
1  &   Henrik Lundqvist  &  G &  928100  &   0.00  & 127.80 &  127.80 & 33.24\\
2   &     Pavel Datsyuk  &   C &  320200  & 103.70  &  15.93  & 119.60 & 21.28 \\
3   &     Henrik Sedin  &   C &  350300  & 100.60  &   0.10 &  100.70 & 13.12 \\
4    &    Alex Ovechkin   &  L &  373500  &  94.81 &   -0.18  &  94.63 & 6.41 \\
5   &    Sidney Crosby  &   C  & 240400  &  98.37 &  -13.32  &  85.06 & 4.66 \\
6  &   Alexander Semin &   L &  271100  &  69.88 &   -0.40  &  69.48 & 3.35 \\
7   &    Evgeni Malkin  &  C &  309400  &  81.26  & -12.63 &   68.63 & 2.47 \\
8   &   Marian Gaborik  &  R &  276700  &  67.61 &   -0.22  &  67.39 & 2.33 \\
9   &   Loui Eriksson  &  L &  330900  &  66.44  &  -0.45 &   65.99 & 2.04 \\
10   &   Jarome Iginla  &  R &  393100  &  73.61  &  -8.72 &   64.89 & 1.02 \\
11   &      Tim Thomas  &  G &  732300  &   0.00  &  63.31  &  63.31 & 0.73 \\
12   &    Joe Thornton  &    C &  360000  &  55.96  &   6.92  &  62.88 & 0.87 \\
13  &   Ilya Kovalchuk   &  L &  376500  &  72.66  & -13.66 &   59.01 & 0.58 \\
14   &   Martin Brodeur   & G &  814100  &   0.00 &   58.64  &  58.64 & 0.29 \\
15   &  Roberto Luongo  &  G &  799600  &   0.00  &  57.10  &  57.10 & 0.73 \\
16  &   Jonathan Toews  &  C &  306800  &  57.22  &  -0.17  &  57.05 & 1.16 \\
17  &   Martin St. Louis  &  R &  384400  &  68.73 &  -12.16 &   56.56 & 0.29 \\
18   &   Jason Spezza  &  C & 318400 &   69.67  & -13.21  &  56.46 & 0.44 \\
19   &   Patrick Sharp   &  R &  300200  &  54.65  &  -1.00  & 53.65 & 0.29 \\
20  & Henrik Zetterberg  & L &  339700  &  52.07  &  -0.80  &  51.27 & 0.43 \\
 & $\cdots$ & & & & $\cdots$& \\
81  &  Zdeno Chara &   D &  436700  & 21.64 &   1.832  & 23.47 & 0 \\
 & $\cdots$ & & & & $\cdots$& \\
\hline
\end{tabular}
\end{center}
\end{table}

\begin{table}
\begin{center}
\caption{The top 10 MVPs and bottom 10 LVPs for the 2011-2012 season, calculated as the rating of a player relative to their team's average and selected by the Lasso method. \label{t:mvp-lvp}}
\begin{tabular}{clcclc}
\hline
Team & MVP & Rel. Rating  & Team &       LVP & Rel. Rating \\
\hline
EDM  &   Jordan Eberle &  0.407       &         N.J   &     Ryan Carter    &  -0.338 \\
T.B  &  Steven Stamkos &  0.334       &         NYI   &  Nino Niederreiter &  -0.315 \\
PIT  &   Sidney Crosby &  0.332       &         DET   &   Tomas Holmstrom  &  -0.266 \\
NYI  &    John Tavares &  0.295       &         BOS   &    Shawn Thornton  &  -0.252 \\
FLA  &   Stephen Weiss &  0.276       &         CHI   &    Michael Frolik  &  -0.238 \\
PHX  &   Adrian Aucoin &  0.221       &         MTL   &     Alexei Emelin  &  -0.229 \\
OTT  &  Marcus Foligno &  0.203       &         T.B   &    Dominic Moore   &  -0.202 \\
WSH  & Alexander Semin &  0.200       &         WSH   &    Michael Knuble  &  -0.200 \\
STL  &    David Perron &  0.200       &         BUF   &      Robyn Regehr  &  -0.178 \\
DAL   &     Jamie Benn &  0.184       &         CGY   &      Tim Jackman   &  -0.173 \\
\hline
\end{tabular}
\end{center}
\end{table}

%Do we want to add more here?

\section{Applications with Variable Selection}\label{s:extra-applications}

Many problems of interest have to do with selecting a relevant subset of predictors from a much larger set. There are several such examples we can carry out with our method that we present here. These methods tend to be considerably faster than operations with Markov Chain Monte Carlo, since we're more concerned with the selection of a subset than in the evaluation of its stochastic properties. A negative consequence of this is that this estimation approach is non-regular, making assessment of uncertainty difficult \citep{dawid1994spbi}. Our primary purpose here is identification, rather than quantification (which is handled well by the full hierarchical Bayesian treatment) and our numerical estimates are presented so that we can compare their magnitudes with effects from the full model.

\subsection{``Most Valuable Player'' Awards, Per Team, Per Season}\label{s:mvp}

The term Most Valuable Player has many interpretations throughout the sports world. One that appeals to us is the notion that a player is most valuable to their team if their team's performance suffers the most compared to a ``replacement'' player in their stead. In the context of this model, we propose that each player should be judged with respect to the rest of their team. Since selecting an exceptional player can be treated as a special case of variable selection, we propose a scheme to pick exceptional players on each team.

We use a model with teams and individual players as predictors. (We omit goaltenders for this ranking due to the confounding with team ratings.) We fix the estimates for team ability and the grand means to be those obtained in Section \ref{s:team-rating}. This is to ensure that all subsequent player ratings obtained will roughly sum to zero, since all ratings are relative to their team rating for each of offense and defense.

We use a single shrinkage penalty for player ratings. Here we choose a single Lasso penalty of $\lambda = 8$ as it produces the highest likelihood for the out-of-sample data in three of five seasons; in the other two, the optimal penalty was such that no player had a non-zero relative rating. In each case, the fit to out-of-sample data was virtually identical for penalties greater than 5. For each team, we select players with the highest and lowest offensive, defensive and overall ratings, and place them in the appropriate MVP and LVP tables. When there are empty cells in the table, we steadily decrease the penalty, filling in empty cells in the MVP and LVP table as new players emerge, and stopping when all cells in the table are filled. (This occurs for between 2 and 5 teams.)

\begin{figure}[!ht]
\begin{center}
\includegraphics[width=\linewidth, natwidth=1000, natheight=500]{\imloc 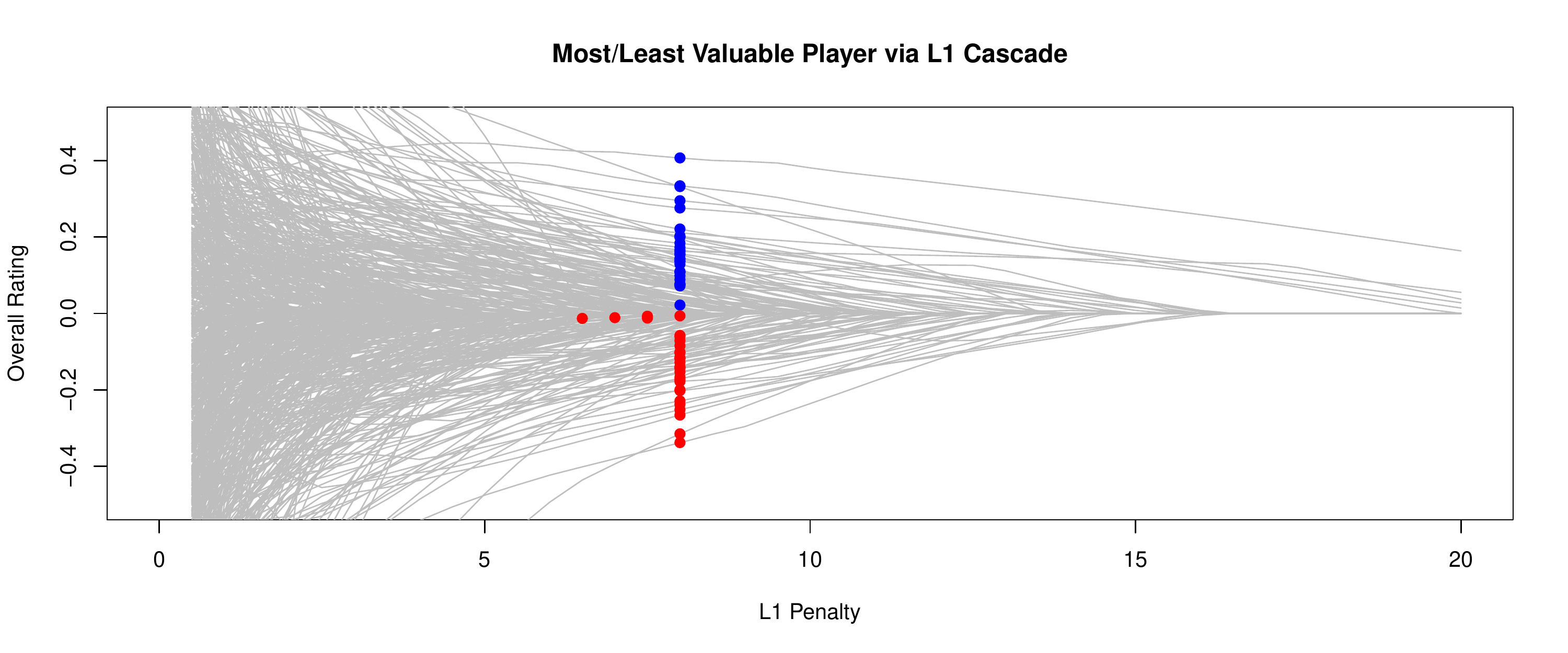}
\end{center}
\caption{The Lasso Cascade method for picking team Most/Least Valuable Players for the 2011-2012 season. Team-level effects are fixed, and player effects are subjected to a steadily decreasing penalty beginning with $\lambda=8$ as chosen by out-of-sample validation. Points indicate where MVPs (in blue) and LVPs (in red) are first declared for overall ability.\label{f:mvp-cascade}}
\end{figure}

Figure \ref{f:mvp-cascade} shows a demonstration of the method for the 2011-2012 season, and Table \ref{t:mvp-lvp} lists the top 10 MVPs and bottom 10 LVPs for that year; a full list of named MVPs and LVPs, for offense, defense and overall, is provided in the supplementary material. Most of the results are consistent with expectations, though we can spot some interesting trends. First, quite often, the most valuable player for offense will be the least valuable player for defense, such as Joffrey Lupul with the 2011-2012 Maple Leafs, or vice versa. In many ways this is not surprising; since the best players have the most ice time, they would be more likely to have ratings that are not shrunk completely to zero on that basis alone, and because these ratings tend to not be correlated (see Figure \ref{f:scatterplot-pos} for ratings in the five-season grand model) it is not surprising that this rating will sometimes be negative.

Second, some of the more surprising Least Valuable Players are centers who specialize in taking faceoffs, often at critical times, such as David Steckel of the Washington Capitals in 2009-2010 and again with the Toronto Maple Leafs in 2011-2012. These players are often brought into the game specifically to take faceoffs, often in their team's defensive zone, before switching off for another player at their next opportunity. Because they are given fewer opportunities to score goals, merely to help prevent them, their offensive ratings will suffer accordingly; their defensive ratings can be insignificant by comparison. Taking puck location into account has been the subject of previous research \citep{thomas2006ippalihs} and its role in this model will be the subject of a future investigation.

\subsection{Identifying Exceptional Player Pair Interactions}\label{s:player-pairs}

If we can select a smaller subset of predictors from a much larger collection, we allow for the possibility of including a substantially large number of extra predictors to any of our models. One compelling inclusion is player interactions; in this context, this would allow us to see whether two players have an additional, detectable ``chemistry'' that yields a higher or lower total in their offensive or defensive abilities. If this is the case, we must see whether there are any corresponding changes to the individual player abilities as well.

Since the MCMC procedure gets considerably slower with the addition of a large number of predictors and coefficients, we use the Lasso method of penalized maximum likelihood to detect a number of non-zero coefficients for the new group. 

We begin by specifying the grand model in Section \ref{s:grand-model}, and we use the mean value of each $\sigma_g$ and $\lambda_g$ as Laplace-Gaussian penalty terms that we will keep fixed for the individual player effects, to allow for and moderate adjustments due to the pair terms.

We then select a subset of player pairs from the database. For this analysis, we took the top 1000 pairs of players in terms of the number of shifts they played together over the five-year period. We use the condition that both players played forward positions or both players played defense, since these groups tend to co-ordinate their play amongst themselves. We add these pairs as predictors to the model. We then estimate the model parameters for a series of Lasso penalty values, labeled $\lambda_{pair}$, on the player-pair terms, in order from strictest to loosest for computational ease. \footnote{We maintain the previously obtained penalty values for player effects.} The choice of penalty term depends on the goal in question; if the goal is to increase predictive accuracy, a penalty term that minimizes out-of-sample error is appropriate.\footnote{If the goal is to select a fixed number of significant partnerships, we would choose the penalty term that yields that count.}

In this case, we find that the penalty $\lambda_{pair} = 8.5$ minimizes the test-set likelihood under cross-validation for these events. Of the 2000 possible parameters to select from (1000 each for $\omega$ and $\delta$), this routine selects 247 non-zero parameters for player pairs for 221 unique player pairs.

\begin{table}
\begin{center}
\caption{The top and bottom five player-pair interactions over 5 NHL seasons. These effects represent the additional total rate beyond the abilities of the players themselves. \label{t:pair-results}}
\begin{tabular}{cccccccc}
\hline
Rank & Player 1 & & Player 2 & & Team & Time(s) & Rating \\
\hline
1 & Brad Boyes & R & Jay McClement & C & STL & 35466 & 0.393 \\
2 & Matt Carle & D & Andrej Meszaros & D & PHI & 41011 & 0.314 \\
3 & Patrice Bergeron & C & Brad Marchand & C & BOS & 85678 & 0.31 \\
4 & Jussi Jokinen & L & Jeff Skinner & C & CAR & 46196 & 0.287 \\
5 & Kris Letang & D & Paul Martin & D & PIT & 40034 & 0.275 \\
\hline
217 & Zach Bogosian & D & John Oduya & D & WPG & 57215 & -0.235 \\
218 & David Booth & L & Michael Santorelli & C & FLA & 34158 & -0.241 \\
219 & Alex Frolov & L & Anze Kopitar & C & LA & 45982 & -0.269 \\
220 & Sidney Crosby & C & Evgeni Malkin & C & PIT & 69217 & -0.283 \\
221 & Ilya Kovalchuk & L & Todd White & C & ATL & 70421 & -0.545 \\
\hline
\end{tabular}
\end{center}
\end{table}

Table \ref{t:pair-results} shows the top and bottom five player pair ratings from the analysis; a more complete list is available in the supporting material. Of particular note is the most extreme case, the pairing of Ilya Kovalchuk of Todd White, whose mutual rating is so low that they effectively wiped out their positive total individual ratings during their time together. Both recorded very high-scoring seasons when they played together, but this accolade effectively masks their mutual liability on defense. The next-lowest pair of Sidney Crosby and Evgeni Malkin is similar; their presence together does not increase their (considerable) offensive prowess beyond their individual levels, but does lead to a substantial increase in the rate of goals scored against their team while they are both on the ice.

Interestingly, the pair of Henrik and Daniel Sedin, twin brothers who play most of their even-strength shifts together, does not appear in the selected group. Indeed, the most total ice time in the top/bottom five is the 135th-most coincident pair of Patrice Bergeron and Brad Marchand from Boston. This suggests that the levels of shrinkage are appropriate for obtaining a reasonable subset of player pairs that have reasonable deviations.

As a final check, the positions of players in the grand rating table are mostly unchanged, so that the original player ratings are reasonably robust to these new additions. Worth noting is that the top two positions in the grand ratings reverse; Sidney Crosby now has the highest player rating over Pavel Datsyuk, due to the removal of the poorer outcomes when he plays with Evgeni Malkin, as opposed to other potential linemates.

%\section{Model Validation, Comparison and Outcomes}\label{s:model-val}

%\subsection{Model Validation}

% Notes added July 15, 2012.

%(**Do the player parameters make sense?  Compare the player parameters to different stats from each season - Goals, Assists, +/-**)
%   Yes. We say this within sections.

%(**Permutation test:  Permute the home/away goal labels, so that goals occur randomly for each team.  Player parameters should all be shrunk to zero when using a reasonable laplace factor.  Does this happen?**)
%   Good question. Not done yet.

%(**Sensitivity analysis:  Take out some goal observations, determine how results change.  Do these changes make sense?**)

%(**Do the player parameter analysis described above, comparing to this season's +/- and next season's +/-.  Also check to see if player ability parameters are consisten across seasons.**)
%   Can't do it, because we don't have it.

%(**Do our player ability parameters predict performance in the following years (measured by points, +/-, or our metrics) better than regular +/-?  Check this.**)
%   Can't do it, because we don't have it.

\section{Discussion and Extensions}\label{s:discussion}

We have presented a model-based method for assessing player ability in ice hockey by treating the game as a competing stochastic process. Given the sheer number of predictors, and the relatively weak explanatory power of each, we use shrinkage methods to improve our estimation of model parameters. We also allow for the possibility of expanding the model specification from a simple flat hazard model to a more general Cox proportional hazards semi-Markov process, to account for other phenomena. In terms of comparisons between players, our method produces similar results for player effects as other approaches \citep{macdonald2011rapsfnp, gramacy2013epchrlr}, suggesting that there is sufficient information in the data to distinguish player ability at a grand level, despite different models. Our method has a key advantage in that it has a specific mechanism for generating hypothetical games, as long as a mechanism for player substitutions is known, and that the physical units of our coefficient estimates correspond directly to a change in the scoring rate.

Here we address potential ways to better extend the model as a useful interpretation of the game. One obvious issue is that the methods for estimating parameters in this model are considerably slower than simple regression, whether we use Monte Carlo methods or functional maximization, especially when more parameters or data points are added. If this method is to ever see conventional and public use, the computation must either be considerably faster, or a new method of estimation must be used. Because this is a highly non-standard likelihood function, it is a complicated matter to improve parameter estimates in a general way. Sequential updating may prove to be the easiest method to improve both methods, particularly with regard to particle filtering for hierarchical Bayesian methods.

We have also assumed that player ability is constant over the period considered, whether this is one season or five. There is considerable reason to expect that player abilities change from year to year in a meaningful way, such as a ``career curve'' \citep{berry1999bdes}, or as simple deviations from a career mean. In this analysis, we chose to use the constant approach for several reasons, mainly that it would grossly magnify the number of parameters in a model where the data is already information-poor. We leave the introduction of single-player variability into this model as a subject of future research.

As a practical matter, there are several factors that can be explored immediately. Many have to do with the use of the time-dependent component of the Cox model, which we have kept as constant and unit-valued to this point.

\subsubsection*{Knowing Location Affects The Short-Term Scoring Rate}

A game of hockey begins with a face-off at center ice, immediately after which neither team is very likely to score in the next few seconds. A distribution for the goal hazard after faceoff was proposed by \citet{thomas2007itgih}, which begins at 0 for both teams and rises to a plateau with an exponential decay. If a team has the puck in their offensive zone, they are more likely to score a goal in the immediate future than the mean rate, and their opponents far less likely.

One approach is to include known puck possession and location terms as covariates in a general model; \citet{macdonald2012egmfentap} in particular uses the zone in which the play starts as a mean-altering covariate. In our case, the natural point to include this is in the time-varying component to the Cox model, by choosing a relative hazard that starts at a rate given the state of play and returns to the overall mean. We expect that this will alleviate the issues highlighted in Section \ref{s:mvp}, wherein some players are frequently substituted in for defensive zone face-offs, a choice that unfairly penalizes their offensive ratings.

One benign side effect of this is that ``garbage goals'' -- those scored after a longer scrum in an offensive zone, taking advantage of continued pressure rather than pure skill -- would be down-weighted, since we would expect a goal to be much more likely in that scenario.

\subsubsection*{Including More Events As Outcomes}

Since a goal is preceded by a shot on goal in the vast majority of cases, one method to improve the modeling framework is to consider shots to be a non-censored terminating state of a model instead of a goal. Since this would lead to a roughly ten-fold count in the number of uncensored events, it would represent a great increase in the precision of estimates, especially if there was no individual variability on what fraction of shots on goal became goals. But this is certainly not the case, since there is significant variety on the fraction of shots that become goals (let alone shots on net) depending on the player; a defenseman's slap shot is considerably less likely than a forward's wrist shot. How we can include this feature in this model framework is an open problem, but may include information on the success rate of shots based on location and type as a post-processing step.

\subsubsection*{Censoring May Be Slightly Informative}

\begin{figure}[ht]
\begin{center}
\includegraphics[width=0.8\linewidth, natwidth=1000, natheight=500]{\imloc 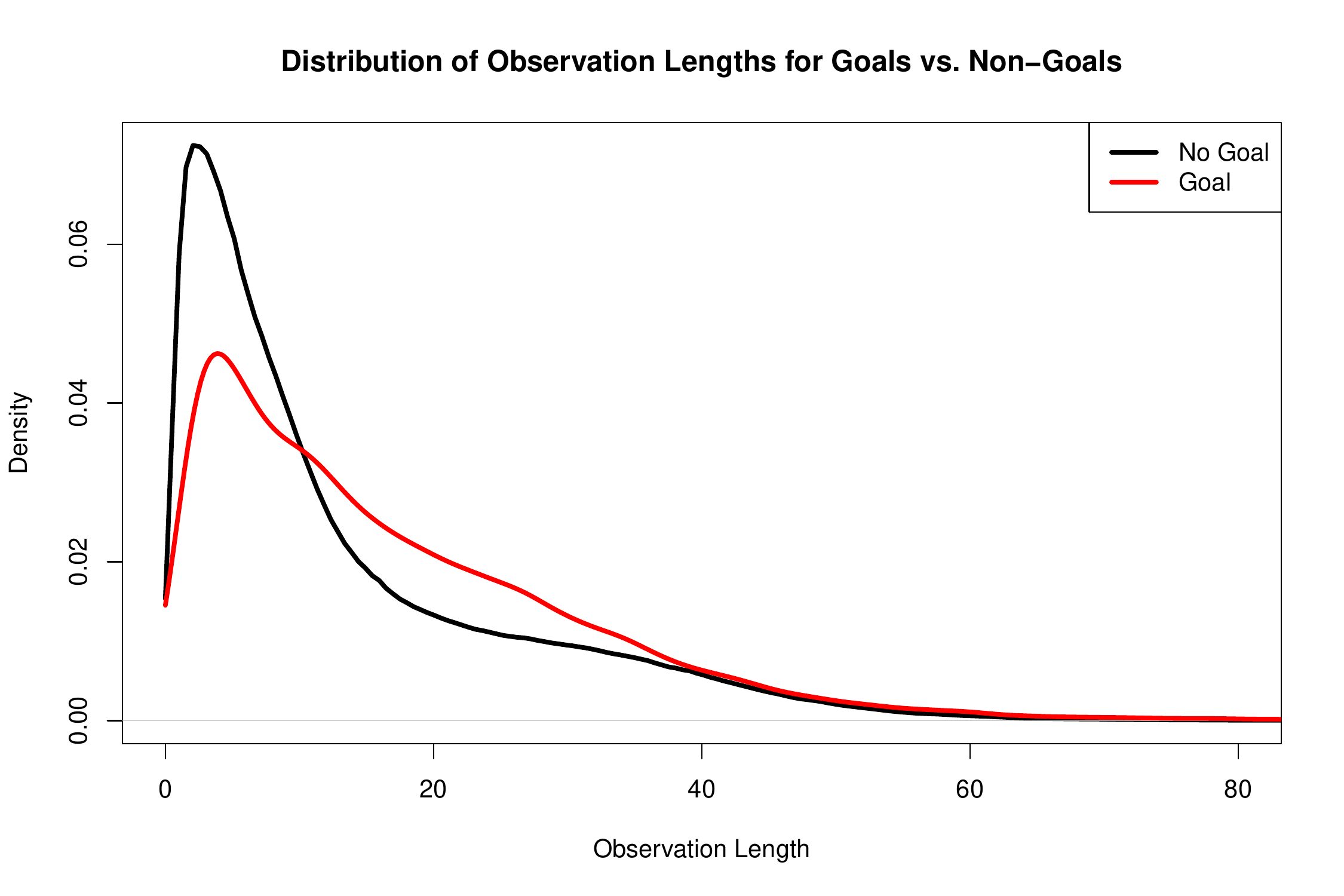}
\end{center}
\caption{The lengths of shifts, conditioned on whether or not a goal was scored to terminate the observation. Shifts that end in goals are slightly longer.\label{f:shifts-goals}}
\end{figure}

Shift lengths are either obtained directly or censored by player changes. One assumptions we make is that the censoring mechanism is roughly exogenous, and does not depend on of influence the state of the game in progress. While this assumption is clearly incorrect, the distributions of shift time are quite similar, as shown in Figure \ref{f:shifts-goals}. Two immediate reasons for this are clear. First, a goal is often scored following a longer scrum in the offensive zone, during which players have no opportunity to change off. Second, the changing process can be sequential; three players change, then shortly after, the other two change off, leading to a bias in short shifts. We expect that this factor can be accounted for, either through modeling or stratification, once we take puck possession and location into account.

\subsubsection*{Does The Power Play Look Like The Process Model?}

When a team has a man-advantage over their opponents, the game tends to look very differently than a smooth stochastic process: the team on the power play sets up shop in their offensive zone, plays keep-away from their opponents and maneuvers to make a shot on goal. The short-handed team's prime goal in this period is not to score, but to remove the danger by clearing the puck from their own zone. (Scoring a short-handed goal is often seen as a bonus rather than the main objective while killing a penalty.)

To extend this model to the power-play situation, we would need to account for this in a principled manner. It may be sufficient to simply change the baseline scoring rates, or to replace the penalized player with an indicator for the power play state, but this is subject to a future investigation and not at all obvious given the apparent differences in game play.

\subsubsection*{Acknowledgements}

{\em We thank Brian Macdonald for introducing us to the shift-based method of dividing the game into components and for fruitful discussions. We also thank an associate editor and two anonymous referees for their helpful comments.}

%\section{Appendix}
%Supplemental material:
%\begin{itemize}
%  \item Fake data test
%  \item How were the parameters calculated?  (Cyclical adjustments to the likelihood)
%  \item Data collection
%\end{itemize}

\bibliographystyle{ims}
\bibliography{references}

\begin{thebibliography}{24}
\expandafter\ifx\csname natexlab\endcsname\relax\def\natexlab#1{#1}\fi
\expandafter\ifx\csname url\endcsname\relax
  \def\url#1{\texttt{#1}}\fi
\expandafter\ifx\csname urlprefix\endcsname\relax\def\urlprefix{URL }\fi
\providecommand{\eprint}[2][]{\url{#2}}

\bibitem[{Beaudoin and Swartz(2010)}]{beaudoin2010sfpgh}
\textsc{Beaudoin, D.} and \textsc{Swartz, T.~B.} (2010).
\newblock {Strategies for Pulling the Goalie in Hockey}.
\newblock \textit{The American Statistician}, \textbf{64}.

\bibitem[{Berry et~al.(1999)Berry, Reese and Larkey}]{berry1999bdes}
\textsc{Berry, S.~M.}, \textsc{Reese, C.~S.} and \textsc{Larkey, P.~D.} (1999).
\newblock {Bridging Different Eras in Sports}.
\newblock \textit{Journal of the American Statistical Association}, \textbf{94}
  661--676.

\bibitem[{Brown(2008)}]{brown2008ipbaftebabm}
\textsc{Brown, L.~D.} (2008).
\newblock {In-season Prediction of Batting Averages -- A Field Test of
  Empirical Bayes and Bayes Methodologies}.
\newblock \textit{The Annals of Applied Statistics}, \textbf{2} 113–152.

\bibitem[{Cook et~al.(2006)Cook, Gelman and Rubin}]{cook2006vsb}
\textsc{Cook, S.~R.}, \textsc{Gelman, A.} and \textsc{Rubin, D.~B.} (2006).
\newblock {Validation of Software for Bayesian Models Using Posterior
  Quantiles}.
\newblock \textit{Journal of Computational and Graphical Statistics},
  \textbf{15} 675--692.

\bibitem[{Cox(1972)}]{cox1972rmald}
\textsc{Cox, D.} (1972).
\newblock {Regression Models and Life-Tables (with discussion)}.
\newblock \textit{Journal of the Royal Statistical Society, Series B},
  \textbf{34} 187--220.

\bibitem[{Dawid(1994)}]{dawid1994spbi}
\textsc{Dawid, A.~P.} (1994).
\newblock {Selection Paradoxes of Bayesian Inference}.
\newblock \textit{Lecture Notes-Monograph Series, Vol. 24, Multivariate
  Analysis and Its Applications} 211--220.

\bibitem[{Gramacy et~al.(2013)Gramacy, Jensen and Taddy}]{gramacy2013epchrlr}
\textsc{Gramacy, R.~B.}, \textsc{Jensen, S.~T.} and \textsc{Taddy, M.} (2013).
\newblock {Estimating Player Contribution in Hockey with Regularized Logistic
  Regression}.
\newblock \\\urlprefix\url{http://arxiv.org/abs/1209.5026}.

\bibitem[{Hirotsu and Wright(2002)}]{hirotsu2002umpmafmdotsatd}
\textsc{Hirotsu, N.} and \textsc{Wright, M.} (2002).
\newblock {Using a Markov Process Model of an Association Football Match to
  Determine the Optimal Timing of Substitution and Tactical Decisions}.
\newblock \textit{Journal of the Operational Research Society}, \textbf{53}.

\bibitem[{Hoerl and Kennard(1970)}]{hoerl1970rrbefnp}
\textsc{Hoerl, A.~E.} and \textsc{Kennard, R.~W.} (1970).
\newblock {Ridge regression: Biased estimation for nonorthogonal problems}.
\newblock \textit{Technometrics}, \textbf{12} 55–67.

\bibitem[{Ilardi and Barzilai(2008)}]{ilardi2008aprnaif2}
\textsc{Ilardi, S.} and \textsc{Barzilai, A.} (2008).
\newblock {Adjusted Plus-Minus Ratings: New and Improved for 2007-2008}.
\newblock \\\urlprefix\url{http://www.82games.com/ilardi2.htm}.

\bibitem[{James and Stein(1961)}]{james1961eql}
\textsc{James, W.} and \textsc{Stein, C.} (1961).
\newblock {Estimation with quadratic loss}.
\newblock \textit{Proc. 4th Berkeley Symp. Probab. Statist}, \textbf{1}
  367--379.

\bibitem[{Lock and Schuckers(2009)}]{lock2009b+rscnp}
\textsc{Lock, D.} and \textsc{Schuckers, M.} (2009).
\newblock {Beyond +/-: A Rating System to Compare NHL Players}.
\newblock Presentation at Joint Statistical Meetings.

\bibitem[{Macdonald(2011)}]{macdonald2011rapsfnp}
\textsc{Macdonald, B.} (2011).
\newblock {A Regression-based Adjusted Plus-Minus Statistic for NHL Players}.
\newblock \textit{Journal of Quantitative Analysis in Sports}, \textbf{7}.

\bibitem[{Macdonald(2012{\natexlab{a}})}]{macdonald2012apfnpurr}
\textsc{Macdonald, B.} (2012{\natexlab{a}}).
\newblock {Adjusted Plus-Minus for NHL Players using Ridge Regression}.
\newblock \\\urlprefix\url{http://arxiv.org/abs/1201.0317v1}.

\bibitem[{Macdonald(2012{\natexlab{b}})}]{macdonald2012egmfentap}
\textsc{Macdonald, B.} (2012{\natexlab{b}}).
\newblock {An Expected Goals Model for Evaluating NHL Teams and Players}.
\newblock In \textit{MIT Sloan Sports Analytics Conference 2012}.

\bibitem[{Morrison(1976)}]{morrison1976otpgpmacsuih}
\textsc{Morrison, D.} (1976).
\newblock {On the Optimal Time to Pull the Goalie: A Poisson Model Applied to a
  Common Strategy Used in Ice Hockey}.
\newblock In \textit{TIMS Studies in Management Science}, vol.~4.

\bibitem[{Rosenbaum(2004)}]{rosenbaum2004mhnphttw}
\textsc{Rosenbaum, D.~T.} (2004).
\newblock {Measuring How NBA Players Help Their Teams Win}.
\newblock \\\urlprefix\url{http://www.82games.com/comm30.htm}.

\bibitem[{Schuckers et~al.(2011)Schuckers, Lock, Wells, Knickerbocker and
  Lock}]{schuckers2011nhlsrbuaoeampaa}
\textsc{Schuckers, M.~E.}, \textsc{Lock, D.~F.}, \textsc{Wells, C.},
  \textsc{Knickerbocker, C.~J.} and \textsc{Lock, R.~H.} (2011).
\newblock {National Hockey League Skater Ratings Based upon All On-Ice Events:
  An Adjusted Minus/Plus Probability (AMPP) Approach}.

\bibitem[{Sill(2010)}]{sill2010ina+uraot}
\textsc{Sill, J.} (2010).
\newblock {Improved NBA Adjusted +/- Using Regularization and Out-of-Sample
  Testing}.
\newblock \textit{MIT Sloan Sports Analytics Conference}.

\bibitem[{Thomas(2006)}]{thomas2006ippalihs}
\textsc{Thomas, A.} (2006).
\newblock {The Impact of Puck Possession and Location on Ice Hockey Strategy}.
\newblock \textit{Journal for Quantitative Analysis in Sports}, \textbf{2}.

\bibitem[{Thomas(2007)}]{thomas2007itgih}
\textsc{Thomas, A.} (2007).
\newblock {Inter-Arrival Times of Goals in Ice Hockey}.
\newblock \textit{Journal of Quantitative Analysis in Sports}, \textbf{3}.

\bibitem[{Tibshirani(1996)}]{tibshirani1996rsasvl}
\textsc{Tibshirani, R.} (1996).
\newblock {Regression Shrinkage and Selection via the Lasso}.
\newblock \textit{Journal of the Royal Statistical Society, Series B
  (Methodology)}, \textbf{58} 267–288.

\bibitem[{Tibshirani(1997)}]{tibshirani1997lmfvscm}
\textsc{Tibshirani, R.} (1997).
\newblock {The Lasso Method for Variable Selection in the Cox Model}.
\newblock \textit{Statistics in Medicine}, \textbf{16} 385--395.

\bibitem[{Zou and Hastie(2005)}]{zou2005regularization}
\textsc{Zou, H.} and \textsc{Hastie, T.} (2005).
\newblock {Regularization and variable selection via the elastic net}.
\newblock \textit{Journal of the Royal Statistical Society: Series B
  (Statistical Methodology)}, \textbf{67} 301--320.

\end{thebibliography}

\newpage
\appendix

{\centering \Large Supplementary Material: Improving NHL Player Ability Ratings with Hazard Function Models for Goal Scoring and Prevention \\ \vskip 0.5cm}

{\begin{center} \large A.C. Thomas, Samuel L. Ventura, Shane Jensen, Stephen Ma \end{center}}

\singlespacing 

\section{Prior Distribution Characteristics}

\vskip 1.0cm
The three prior families we consider (L1, L2, L1+L2) have slightly different properties, as shown in this figure: 

\begin{center}
\includegraphics[width=\linewidth]{\imloc 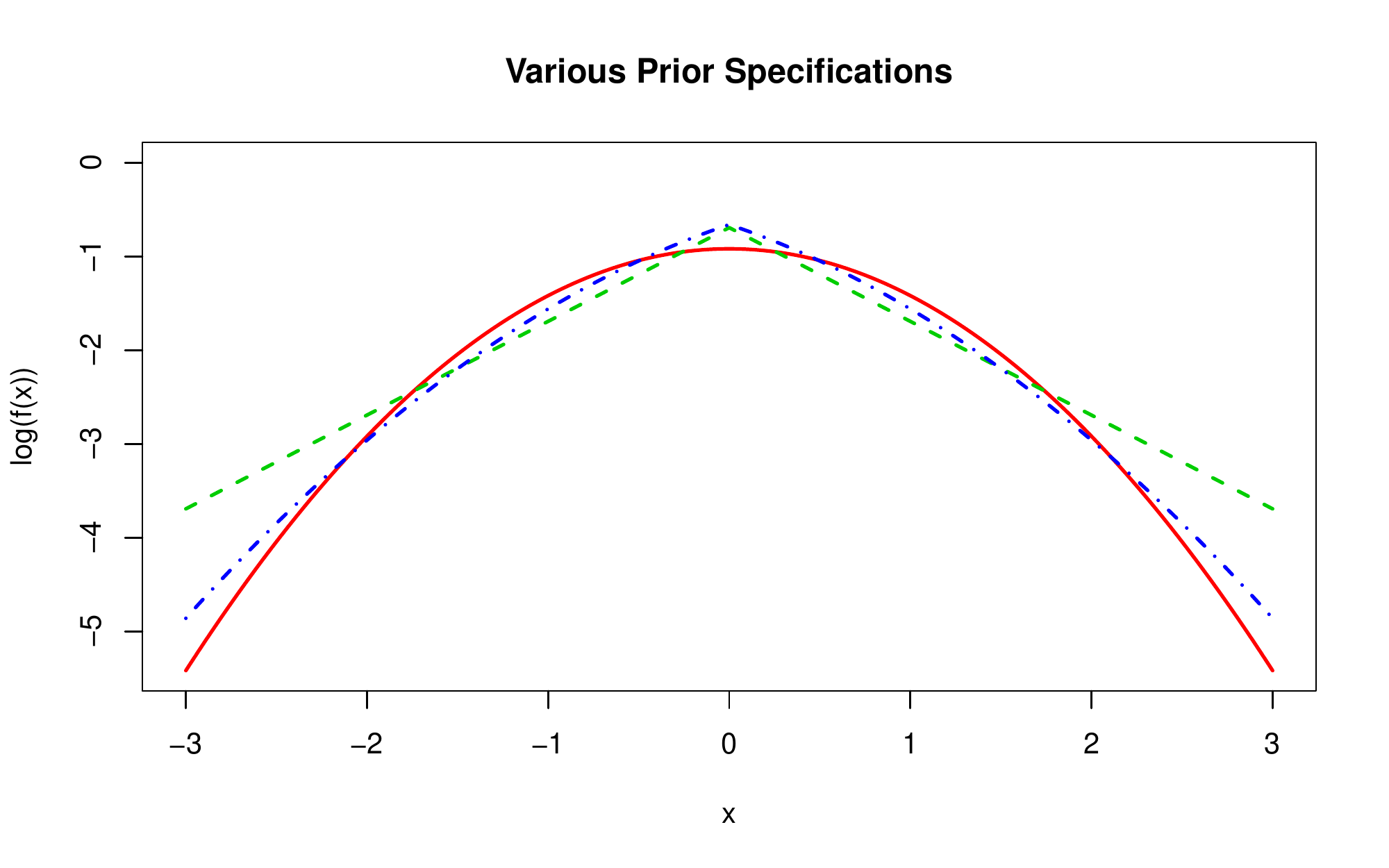}
\end{center}
Each line represents the log-probability density for three families for the prior/penalty distribution on player parameters. Red/solid is the Gaussian, which is smoothly varying with light tails; green/dash is the Laplace, which is sharp at zero, with exponential tails; and blue/dot-dash is the Laplace-Gaussian, which is also sharp with exponential tails, and has one additional parameter to compromise between the other two. All three distributions have unit variance.

\section{Player Abilities Across All Five Seasons}

\small

Goaltenders:\\
\begin{tabular}{lcccccc}
\hline
Rank & Player & $\omega_i - \delta_i$ & $\Delta(\omega_i - \delta_i)$ & Time (s) & $\omega_i$ & $\delta_i$ \\
\hline
1 & HENRIK.LUNDQVIST & 0.186 & 0.0546 & 928000 & 0 & -0.186 \\
2 & TIM.THOMAS & 0.12 & 0.0581 & 732000 & 0 & -0.12 \\
3 & JONATHAN.QUICK & 0.102 & 0.0594 & 662000 & 0 & -0.102 \\
4 & MARTIN.BRODEUR & 0.101 & 0.0571 & 814000 & 0 & -0.101 \\
5 & ROBERTO.LUONGO & 0.1 & 0.0567 & 8e+05 & 0 & -0.1 \\
6 & PEKKA.RINNE & 0.0917 & 0.0556 & 678000 & 0 & -0.0917 \\
7 & CORY.SCHNEIDER & 0.0753 & 0.0761 & 169000 & 0 & -0.0753 \\
8 & DOMINIK.HASEK & 0.0715 & 0.0815 & 102000 & 0 & -0.0715 \\
9 & ANTTI.NIEMI & 0.0602 & 0.0587 & 469000 & 0 & -0.0602 \\
10 & ILJA.BRYZGALOV & 0.0592 & 0.0484 & 859000 & 0 & -0.0592 \\
11 & SEMYON.VARLAMOV & 0.056 & 0.0635 & 308000 & 0 & -0.056 \\
12 & MIKE.SMITH & 0.0509 & 0.0517 & 539000 & 0 & -0.0509 \\
13 & KARI.LEHTONEN & 0.0502 & 0.0473 & 617000 & 0 & -0.0502 \\
14 & EVGENI.NABOKOV & 0.0469 & 0.0496 & 674000 & 0 & -0.0469 \\
15 & ERIK.ERSBERG & 0.0455 & 0.069 & 125000 & 0 & -0.0455 \\
\hline
145 & J-SEBASTIEN.AUBIN & -0.052 & 0.078 & 36900 & 0 & 0.052 \\
146 & JUSTIN.POGGE & -0.0552 & 0.0801 & 17400 & 0 & 0.0552 \\
147 & PATRICK.LALIME & -0.0637 & 0.0626 & 188000 & 0 & 0.0637 \\
148 & HANNU.TOIVONEN & -0.0665 & 0.0791 & 52500 & 0 & 0.0665 \\
149 & ANDREW.RAYCROFT & -0.106 & 0.071 & 231000 & 0 & 0.106 \\
\hline
\end{tabular}
\vskip 1cm

Wingers:\\
\begin{tabular}{lcccccc}
\hline
Rank & Player & $\omega_i - \delta_i$ & $\Delta(\omega_i - \delta_i)$ & Time (s) & $\omega_i$ & $\delta_i$ \\
\hline
1 & ALEXANDER.SEMIN & 0.321 & 0.0744 & 271000 & 0.323 & 0.00221 \\
2 & ALEX.OVECHKIN & 0.318 & 0.0805 & 374000 & 0.319 & 0.000724 \\
3 & MARIAN.GABORIK & 0.308 & 0.0875 & 277000 & 0.309 & 0.00118 \\
4 & LOUI.ERIKSSON & 0.258 & 0.0814 & 331000 & 0.26 & 0.00202 \\
5 & ALEXANDER.RADULOV & 0.249 & 0.127 & 71600 & 0.265 & 0.0161 \\
6 & PATRICK.SHARP & 0.234 & 0.0833 & 3e+05 & 0.239 & 0.00494 \\
7 & ALEX.TANGUAY & 0.232 & 0.0795 & 278000 & 0.235 & 0.00253 \\
8 & RADIM.VRBATA & 0.23 & 0.109 & 249000 & 0.187 & -0.0432 \\
9 & JAKUB.VORACEK & 0.227 & 0.0905 & 230000 & 0.239 & 0.012 \\
10 & BOBBY.RYAN & 0.221 & 0.0928 & 284000 & 0.232 & 0.0114 \\
11 & THOMAS.VANEK & 0.22 & 0.0915 & 290000 & 0.241 & 0.0203 \\
12 & JAROME.IGINLA & 0.211 & 0.0898 & 393000 & 0.245 & 0.0334 \\
13 & ZACH.PARISE & 0.206 & 0.0848 & 295000 & 0.202 & -0.0043 \\
14 & HENRIK.ZETTERBERG & 0.201 & 0.0831 & 340000 & 0.205 & 0.00349 \\
15 & SCOTT.HARTNELL & 0.199 & 0.0851 & 322000 & 0.205 & 0.00604 \\
\hline
510 & COLTON.ORR & -0.253 & 0.139 & 114000 & -0.27 & -0.0178 \\
511 & RYAN.HOLLWEG & -0.267 & 0.165 & 44400 & -0.268 & -0.00148 \\
512 & STEPHANE.VEILLEUX & -0.267 & 0.114 & 174000 & -0.266 & 0.00117 \\
513 & NINO.NIEDERREITER & -0.281 & 0.171 & 38800 & -0.26 & 0.0211 \\
514 & RAITIS.IVANANS & -0.292 & 0.151 & 80100 & -0.297 & -0.00487 \\
\hline
\end{tabular}
\vskip 1cm

\newpage

Centers:\\
\begin{tabular}{lcccccc}
\hline
Rank & Player & $\omega_i - \delta_i$ & $\Delta(\omega_i - \delta_i)$ & Time (s) & $\omega_i$ & $\delta_i$ \\
\hline
1 & PAVEL.DATSYUK & 0.463 & 0.105 & 320000 & 0.392 & -0.0711 \\
2 & SIDNEY.CROSBY & 0.388 & 0.116 & 240000 & 0.474 & 0.0855 \\
3 & HENRIK.SEDIN & 0.355 & 0.133 & 350000 & 0.354 & -0.000419 \\
4 & PATRICE.BERGERON & 0.28 & 0.112 & 239000 & 0.256 & -0.0246 \\
5 & EVGENI.MALKIN & 0.266 & 0.0907 & 309000 & 0.328 & 0.0623 \\
6 & JONATHAN.TOEWS & 0.243 & 0.0978 & 307000 & 0.244 & 0.000816 \\
7 & JOE.THORNTON & 0.235 & 0.0974 & 360000 & 0.207 & -0.028 \\
8 & JASON.SPEZZA & 0.217 & 0.108 & 318000 & 0.281 & 0.0633 \\
9 & NATHAN.HORTON & 0.207 & 0.103 & 286000 & 0.259 & 0.0521 \\
10 & MATS.SUNDIN & 0.195 & 0.132 & 91200 & 0.195 & -2.97e-05 \\
11 & JORDAN.EBERLE & 0.185 & 0.134 & 124000 & 0.21 & 0.0252 \\
12 & STEPHEN.WEISS & 0.181 & 0.0966 & 328000 & 0.194 & 0.013 \\
13 & JEFF.CARTER & 0.179 & 0.0857 & 307000 & 0.186 & 0.00652 \\
14 & ALEXANDER.STEEN & 0.179 & 0.0994 & 267000 & 0.124 & -0.0551 \\
15 & MARC.SAVARD & 0.175 & 0.114 & 181000 & 0.178 & 0.00354 \\
\hline
416 & NICK.SPALING & -0.156 & 0.131 & 126000 & -0.152 & 0.0039 \\
417 & COLTON.GILLIES & -0.16 & 0.154 & 67000 & -0.171 & -0.011 \\
418 & TOM.PYATT & -0.17 & 0.136 & 107000 & -0.164 & 0.00574 \\
419 & RADEK.BONK & -0.19 & 0.141 & 106000 & -0.181 & 0.00978 \\
420 & ROD.PELLEY & -0.305 & 0.177 & 121000 & -0.334 & -0.0288 \\
\hline
\end{tabular}
\vskip 1cm

Defensemen: \\
\begin{tabular}{lcccccc}
\hline
Rank & Player & $\omega_i - \delta_i$ & $\Delta(\omega_i - \delta_i)$ & Time (s) & $\omega_i$ & $\delta_i$ \\
\hline
1 & ZDENO.CHARA & 0.077 & 0.0739 & 437000 & 0.0708 & -0.00619 \\
2 & MARK.STREIT & 0.0427 & 0.0626 & 303000 & 0.0485 & 0.00585 \\
3 & JAROSLAV.SPACEK & 0.0373 & 0.0528 & 297000 & 0.0374 & 0.000195 \\
4 & MIKE.GREEN & 0.0355 & 0.0527 & 315000 & 0.0374 & 0.00192 \\
5 & MATT.CARLE & 0.0341 & 0.0454 & 379000 & 0.0334 & -0.000666 \\
6 & DAN.HAMHUIS & 0.0335 & 0.0462 & 393000 & 0.0333 & -0.000192 \\
7 & IAN.WHITE & 0.0325 & 0.0471 & 395000 & 0.0341 & 0.00157 \\
8 & TOM.GILBERT & 0.0309 & 0.0452 & 387000 & 0.0321 & 0.00119 \\
9 & FILIP.KUBA & 0.0261 & 0.0461 & 331000 & 0.0284 & 0.00231 \\
10 & LUBOMIR.VISNOVSKY & 0.026 & 0.0433 & 362000 & 0.0284 & 0.00245 \\
11 & KRIS.LETANG & 0.025 & 0.0443 & 326000 & 0.027 & 0.002 \\
12 & BRENT.BURNS & 0.022 & 0.0427 & 353000 & 0.0212 & -0.000733 \\
13 & NICKLAS.LIDSTROM & 0.0215 & 0.0388 & 391000 & 0.0117 & -0.00982 \\
14 & ALEX.GOLIGOSKI & 0.0206 & 0.0429 & 253000 & 0.0238 & 0.00326 \\
15 & KENT.HUSKINS & 0.0202 & 0.0452 & 225000 & 0.014 & -0.00616 \\
\hline
504 & NICLAS.WALLIN & -0.0266 & 0.048 & 225000 & -0.0242 & 0.00235 \\
505 & GARNET.EXELBY & -0.0283 & 0.0499 & 151000 & -0.0261 & 0.00218 \\
506 & LUCA.SBISA & -0.0295 & 0.0518 & 167000 & -0.0285 & 0.000981 \\
507 & ANTON.VOLCHENKOV & -0.0296 & 0.0559 & 303000 & -0.0342 & -0.00456 \\
508 & FRANCOIS.BEAUCHEMIN & -0.0315 & 0.045 & 373000 & -0.0299 & 0.00161 \\
\hline
\end{tabular}

\newpage

\section{Team MVP/LVP By Season}
 \scriptsize
2007-2008: \\

\begin{tabular}{lccc}
\hline
Team & MVP Offense & MVP Defense & MVP Total \\
\hline 
ANA & RYAN.GETZLAF & KENT.HUSKINS & RYAN.GETZLAF \\
ATL & ILYA.KOVALCHUK & BRYAN.LITTLE & ILYA.KOVALCHUK \\
BOS & ZDENO.CHARA & AARON.WARD & ZDENO.CHARA \\
BUF & JAROSLAV.SPACEK & ALES.KOTALIK & JAROSLAV.SPACEK \\
CAR & BRET.HEDICAN & GLEN.WESLEY & BRET.HEDICAN \\
CBJ & JAN.HEJDA & JAN.HEJDA & JAN.HEJDA \\
CGY & JAROME.IGINLA & ADRIAN.AUCOIN & JAROME.IGINLA \\
CHI & JONATHAN.TOEWS & DUNCAN.KEITH & JONATHAN.TOEWS \\
COL & PAUL.STASTNY & KURT.SAUER & PAUL.STASTNY \\
DAL & BRENDEN.MORROW & STEVE.OTT & BRENDEN.MORROW \\
DET & PAVEL.DATSYUK & NICKLAS.LIDSTROM & PAVEL.DATSYUK \\
EDM & JONI.PITKANEN & MATT.GREENE & MATT.GREENE \\
FLA & NATHAN.HORTON & JASSEN.CULLIMORE & NATHAN.HORTON \\
L.A & ALEX.FROLOV & KEVIN.DALLMAN & KEVIN.DALLMAN \\
MIN & MARIAN.GABORIK & JAMES.SHEPPARD & MARIAN.GABORIK \\
MTL & ANDREI.KASTSITSYN & FRANCIS.BOUILLON & ANDREI.KASTSITSYN \\
N.J & DAVID.ODUYA & DAVID.ODUYA & DAVID.ODUYA \\
NSH & JASON.ARNOTT & JERRED.SMITHSON & JASON.ARNOTT \\
NYI & MIKE.COMRIE & JOSEF.VASICEK & JOSEF.VASICEK \\
NYR & SEAN.AVERY & MAREK.MALIK & SEAN.AVERY \\
OTT & DANY.HEATLEY & CHRIS.PHILLIPS & DANY.HEATLEY \\
PHI & BRAYDON.COBURN & BRAYDON.COBURN & BRAYDON.COBURN \\
PHX & SHANE.DOAN & ZBYNEK.MICHALEK & SHANE.DOAN \\
PIT & EVGENI.MALKIN & JORDAN.STAAL & EVGENI.MALKIN \\
S.J & JOE.THORNTON & JONATHAN.CHEECHOO & JOE.THORNTON \\
STL & KEITH.TKACHUK & DAVID.PERRON & DAVID.PERRON \\
T.B & VINCENT.LECAVALIER & MICHEL.OUELLET & MICHEL.OUELLET \\
TOR & MATS.SUNDIN & BRYAN.MCCABE & MATS.SUNDIN \\
VAN & MARKUS.NASLUND & SAMI.SALO & MARKUS.NASLUND \\
WSH & ALEX.OVECHKIN & BOYD.GORDON & ALEX.OVECHKIN \\
\hline
\end{tabular}

%\newpage
%2007-2008: \\

\begin{tabular}{lccc}
\hline
Team & LVP Offense & LVP Defense & LVP Total \\
\hline 
ANA & TRAVIS.MOEN & FRANCOIS.BEAUCHEMIN & TRAVIS.MOEN \\
ATL & STEVE.MCCARTHY & ILYA.KOVALCHUK & STEVE.MCCARTHY \\
BOS & SHANE.HNIDY & PHIL.KESSEL & SHANE.HNIDY \\
BUF & NOLAN.PRATT & THOMAS.VANEK & NOLAN.PRATT \\
CAR & NICLAS.WALLIN & ERIC.STAAL & NICLAS.WALLIN \\
CBJ & DAVID.VYBORNY & RICK.NASH & RICK.NASH \\
CGY & STEPHANE.YELLE & ANDERS.ERIKSSON & STEPHANE.YELLE \\
CHI & CRAIG.ADAMS & CRAIG.ADAMS & CRAIG.ADAMS \\
COL & KARLIS.SKRASTINS & RYAN.SMYTH & KARLIS.SKRASTINS \\
DAL & MATTIAS.NORSTROM & STEPHANE.ROBIDAS & STEPHANE.ROBIDAS \\
DET & DALLAS.DRAKE & ANDREAS.LILJA & DALLAS.DRAKE \\
EDM & JARRET.STOLL & SAM.GAGNER & JARRET.STOLL \\
FLA & BRANISLAV.MEZEI & OLLI.JOKINEN & BRANISLAV.MEZEI \\
L.A & MICHAL.HANDZUS & PATRICK.OSULLIVAN & MICHAL.HANDZUS \\
MIN & BRIAN.ROLSTON & SEAN.HILL & SEAN.HILL \\
MTL & JOSH.GORGES & ANDREI.MARKOV & JOSH.GORGES \\
N.J & SERGEI.BRYLIN & VITALY.VISHNEVSKI & SERGEI.BRYLIN \\
NSH & JERRED.SMITHSON & ALEXANDER.RADULOV & JERRED.SMITHSON \\
NYI & RADEK.MARTINEK & MIKE.COMRIE & MIKE.COMRIE \\
NYR & RYAN.HOLLWEG & CHRIS.DRURY & RYAN.HOLLWEG \\
OTT & ANTON.VOLCHENKOV & WADE.REDDEN & ANTON.VOLCHENKOV \\
PHI & SAMI.KAPANEN & DANIEL.BRIERE & SAMI.KAPANEN \\
PHX & MICHAEL.YORK & SHANE.DOAN & MICHAEL.YORK \\
PIT & ADAM.HALL & RYAN.WHITNEY & RYAN.WHITNEY \\
S.J & MARC-EDOUARD.VLASIC & SANDIS.OZOLINSH & SANDIS.OZOLINSH \\
STL & RYAN.JOHNSON & PAUL.KARIYA & RYAN.JOHNSON \\
T.B & NICK.TARNASKY & DAN.BOYLE & NICK.TARNASKY \\
TOR & TOMAS.KABERLE & JIRI.TLUSTY & TOMAS.KABERLE \\
VAN & SAMI.SALO & AARON.MILLER & BYRON.RITCHIE \\
WSH & SHAONE.MORRISONN & MICHAEL.NYLANDER & MICHAEL.NYLANDER \\
\hline
\end{tabular}

\newpage
2008-2009: \\

\begin{tabular}{lccc}
\hline
Team & MVP Offense & MVP Defense & MVP Total \\
\hline
ANA & COREY.PERRY & BRETT.FESTERLING & COREY.PERRY \\
ATL & ZACH.BOGOSIAN & MARTY.REASONER & ZACH.BOGOSIAN \\
BOS & MARC.SAVARD & DAVID.KREJCI & MARC.SAVARD \\
BUF & THOMAS.VANEK & ADAM.MAIR & THOMAS.VANEK \\
CAR & ERIC.STAAL & PATRICK.EAVES & ERIC.STAAL \\
CBJ & JAKUB.VORACEK & CHRISTIAN.BACKMAN & JAKUB.VORACEK \\
CGY & MATTHEW.LOMBARDI & CORY.SARICH & CORY.SARICH \\
CHI & ANDREW.LADD & DUSTIN.BYFUGLIEN & ANDREW.LADD \\
COL & RUSLAN.SALEI & IAN.LAPERRIERE & IAN.LAPERRIERE \\
DAL & LOUI.ERIKSSON & STEPHANE.ROBIDAS & LOUI.ERIKSSON \\
DET & PAVEL.DATSYUK & BRETT.LEBDA & PAVEL.DATSYUK \\
EDM & DENIS.GREBESHKOV & LUBOMIR.VISNOVSKY & DENIS.GREBESHKOV \\
FLA & STEPHEN.WEISS & KARLIS.SKRASTINS & STEPHEN.WEISS \\
L.A & KYLE.QUINCEY & SEAN.ODONNELL & SEAN.ODONNELL \\
MIN & ANDREW.BRUNETTE & ANTTI.MIETTINEN & ANDREW.BRUNETTE \\
MTL & JOSH.GORGES & MAXIM.LAPIERRE & JOSH.GORGES \\
N.J & ZACH.PARISE & MIKE.MOTTAU & ZACH.PARISE \\
NSH & JASON.ARNOTT & ANTTI.PIHLSTROM & JASON.ARNOTT \\
NYI & MARK.STREIT & SEAN.BERGENHEIM & MARK.STREIT \\
NYR & NIKOLAI.ZHERDEV & RYAN.CALLAHAN & NIKOLAI.ZHERDEV \\
OTT & DANIEL.ALFREDSSON & FILIP.KUBA & DANIEL.ALFREDSSON \\
PHI & JEFF.CARTER & CLAUDE.GIROUX & JEFF.CARTER \\
PHX & STEVE.REINPRECHT & KEN.KLEE & STEVE.REINPRECHT \\
PIT & EVGENI.MALKIN & ROB.SCUDERI & EVGENI.MALKIN \\
S.J & DEVIN.SETOGUCHI & MICHAEL.GRIER & MICHAEL.GRIER \\
STL & PATRIK.BERGLUND & PATRIK.BERGLUND & PATRIK.BERGLUND \\
T.B & MARTIN.ST LOUIS & ADAM.HALL & MARTIN.ST LOUIS \\
TOR & ALEXEI.PONIKAROVSKY & IAN.WHITE & ALEXEI.PONIKAROVSKY \\
VAN & HENRIK.SEDIN & DANIEL.SEDIN & HENRIK.SEDIN \\
WSH & ALEX.OVECHKIN & MILAN.JURCINA & ALEX.OVECHKIN \\
\hline
\end{tabular}

%\newpage
%2008-2009: \\

\begin{tabular}{lccc}
\hline
Team & LVP Offense & LVP Defense & LVP Total \\
\hline
ANA & TRAVIS.MOEN & ROB.NIEDERMAYER & TRAVIS.MOEN \\
ATL & GARNET.EXELBY & ILYA.KOVALCHUK & GARNET.EXELBY \\
BOS & SHAWN.THORNTON & VLADIMIR.SOBOTKA & SHAWN.THORNTON \\
BUF & JOCHEN.HECHT & THOMAS.VANEK & JOCHEN.HECHT \\
CAR & ROD.BRINDAMOUR & JOE.CORVO & ROD.BRINDAMOUR \\
CBJ & ANDREW.MURRAY & KRISTIAN.HUSELIUS & KRISTIAN.HUSELIUS \\
CGY & ERIC.NYSTROM & DION.PHANEUF & ERIC.NYSTROM \\
CHI & BEN.EAGER & BRIAN.CAMPBELL & BRIAN.CAMPBELL \\
COL & DARCY.TUCKER & JORDAN.LEOPOLD & JORDAN.LEOPOLD \\
DAL & JOEL.LUNDQVIST & MATT.NISKANEN & JOEL.LUNDQVIST \\
DET & KRIS.DRAPER & DAN.CLEARY & KRIS.DRAPER \\
EDM & JASON.STRUDWICK & LIAM.REDDOX & JASON.STRUDWICK \\
FLA & JAY.BOUWMEESTER & NATHAN.HORTON & JAY.BOUWMEESTER \\
L.A & RAITIS.IVANANS & ANZE.KOPITAR & RAITIS.IVANANS \\
MIN & STEPHANE.VEILLEUX & MARTIN.SKOULA & STEPHANE.VEILLEUX \\
MTL & ALEX.KOVALEV & ANDREI.KASTSITSYN & ALEX.KOVALEV \\
N.J & JAY.PANDOLFO & COLIN.WHITE & JAY.PANDOLFO \\
NSH & ANTTI.PIHLSTROM & JASON.ARNOTT & RADEK.BONK \\
NYI & THOMAS.POCK & FREDDY.MEYER & THOMAS.POCK \\
NYR & COLTON.ORR & MARC.STAAL & COLTON.ORR \\
OTT & ANTON.VOLCHENKOV & JASON.SPEZZA & ANTON.VOLCHENKOV \\
PHI & ANDREAS.NODL & ARRON.ASHAM & ANDREAS.NODL \\
PHX & DAN.CARCILLO & DAVID.HALE & DAVID.HALE \\
PIT & MARK.EATON & KRIS.LETANG & MARK.EATON \\
S.J & JODY.SHELLEY & DEVIN.SETOGUCHI & DEVIN.SETOGUCHI \\
STL & MIKE.WEAVER & BARRET.JACKMAN & BARRET.JACKMAN \\
T.B & ADAM.HALL & MARK.RECCHI & MARK.RECCHI \\
TOR & JOHN.MITCHELL & NIK.ANTROPOV & JOHN.MITCHELL \\
VAN & MASON.RAYMOND & KEVIN.BIEKSA & MASON.RAYMOND \\
WSH & MILAN.JURCINA & TOMAS.FLEISCHMANN & DONALD.BRASHEAR \\
\hline
\end{tabular}

%\newpage
%2009-2010: \\

\begin{tabular}{lccc}
\hline
Team & MVP Offense & MVP Defense & MVP Total \\
\hline
ANA & RYAN.GETZLAF & GEORGE.PARROS & GEORGE.PARROS \\
ATL & ILYA.KOVALCHUK & CHRIS.THORBURN & ILYA.KOVALCHUK \\
BOS & ZDENO.CHARA & MARCO.STURM & ZDENO.CHARA \\
BUF & JOCHEN.HECHT & TONI.LYDMAN & TONI.LYDMAN \\
CAR & ERIC.STAAL & BRETT.CARSON & ERIC.STAAL \\
CBJ & JAKUB.VORACEK & NATHAN.PAETSCH & NATHAN.PAETSCH \\
CGY & RENE.BOURQUE & CURTIS.GLENCROSS & CURTIS.GLENCROSS \\
CHI & PATRICK.SHARP & PATRICK.SHARP & PATRICK.SHARP \\
COL & CHRIS.STEWART & DARCY.TUCKER & WOJTEK.WOLSKI \\
DAL & BRAD.RICHARDS & MARK.FISTRIC & MARK.FISTRIC \\
DET & HENRIK.ZETTERBERG & DREW.MILLER & DREW.MILLER \\
EDM & DUSTIN.PENNER & TOM.GILBERT & DUSTIN.PENNER \\
FLA & NATHAN.HORTON & DENNIS.SEIDENBERG & NATHAN.HORTON \\
L.A & WAYNE.SIMMONDS & DREW.DOUGHTY & WAYNE.SIMMONDS \\
MIN & KIM.JOHNSSON & ERIC.BELANGER & KIM.JOHNSSON \\
MTL & BRIAN.GIONTA & JOSH.GORGES & BRIAN.GIONTA \\
N.J & ZACH.PARISE & MARK.FRASER & ZACH.PARISE \\
NSH & DAN.HAMHUIS & FRANCIS.BOUILLON & FRANCIS.BOUILLON \\
NYI & MARK.STREIT & ANDREW.MACDONALD & MARK.STREIT \\
NYR & MARIAN.GABORIK & MARC.STAAL & MARIAN.GABORIK \\
OTT & ALEXANDRE.PICARD & MARCUS.FOLIGNO & MARCUS.FOLIGNO \\
PHI & MATT.CARLE & DAN.CARCILLO & MATT.CARLE \\
PHX & ZBYNEK.MICHALEK & TAYLOR.PYATT & TAYLOR.PYATT \\
PIT & SIDNEY.CROSBY & JORDAN.STAAL & SIDNEY.CROSBY \\
S.J & PATRICK.MARLEAU & MANNY.MALHOTRA & MANNY.MALHOTRA \\
STL & ALEXANDER.STEEN & BRAD.WINCHESTER & BRAD.WINCHESTER \\
T.B & STEVEN.STAMKOS & STEPHANE.VEILLEUX & VICTOR.HEDMAN \\
TOR & IAN.WHITE & COLTON.ORR & IAN.WHITE \\
VAN & DANIEL.SEDIN & KYLE.WELLWOOD & DANIEL.SEDIN \\
WSH & ALEX.OVECHKIN & JEFF.SCHULTZ & ALEX.OVECHKIN \\
\hline
\end{tabular}

%\newpage
%2009-2010: \\

\begin{tabular}{lccc}
\hline
Team & LVP Offense & LVP Defense & LVP Total \\
\hline
ANA & BRAD.MARCHAND & COREY.PERRY & BRAD.MARCHAND \\
ATL & ZACH.BOGOSIAN & MAXIM.AFINOGENOV & MAXIM.AFINOGENOV \\
BOS & STEVE.BEGIN & BLAKE.WHEELER & BLAKE.WHEELER \\
BUF & CRAIG.RIVET & CLARKE.MACARTHUR & CRAIG.RIVET \\
CAR & ROD.BRINDAMOUR & ROD.BRINDAMOUR & ROD.BRINDAMOUR \\
CBJ & JARED.BOLL & R.J..UMBERGER & JARED.BOLL \\
CGY & JAY.BOUWMEESTER & RENE.BOURQUE & JAY.BOUWMEESTER \\
CHI & DUSTIN.BYFUGLIEN & ANDREW.LADD & DUSTIN.BYFUGLIEN \\
COL & DARCY.TUCKER & CHRIS.STEWART & DARCY.TUCKER \\
DAL & MATT.NISKANEN & BRAD.RICHARDS & MATT.NISKANEN \\
DET & KIRK.MALTBY & JONATHAN.ERICSSON & KIRK.MALTBY \\
EDM & TAYLOR.CHORNEY & PATRICK.OSULLIVAN & PATRICK.OSULLIVAN \\
FLA & DENNIS.SEIDENBERG & NATHAN.HORTON & KEITH.BALLARD \\
L.A & RAITIS.IVANANS & JACK.JOHNSON & JACK.JOHNSON \\
MIN & CAL.CLUTTERBUCK & MARTIN.HAVLAT & CAL.CLUTTERBUCK \\
MTL & MAXIM.LAPIERRE & PAUL.MARA & MAXIM.LAPIERRE \\
N.J & NICKLAS.BERGFORS & MIKE.MOTTAU & NICKLAS.BERGFORS \\
NSH & FRANCIS.BOUILLON & MARTIN.ERAT & MARTIN.ERAT \\
NYI & NATE.THOMPSON & KYLE.OKPOSO & NATE.THOMPSON \\
NYR & CHRIS.HIGGINS & MICHAEL.DEL ZOTTO & CHRIS.HIGGINS \\
OTT & RYAN.SHANNON & ALEX.KOVALEV & ALEX.KOVALEV \\
PHI & RYAN.PARENT & OSKARS.BARTULIS & OSKARS.BARTULIS \\
PHX & LAURI.KORPIKOSKI & ED.JOVANOVSKI & LAURI.KORPIKOSKI \\
PIT & RUSLAN.FEDOTENKO & EVGENI.MALKIN & RUSLAN.FEDOTENKO \\
S.J & JED.ORTMEYER & DAN.BOYLE & JED.ORTMEYER \\
STL & B.J..CROMBEEN & ERIC.BREWER & ERIC.BREWER \\
T.B & STEPHANE.VEILLEUX & STEVEN.STAMKOS & STEPHANE.VEILLEUX \\
TOR & FRANCOIS.BEAUCHEMIN & MATT.STAJAN & MATT.STAJAN \\
VAN & KYLE.WELLWOOD & RYAN.KESLER & RYAN.KESLER \\
WSH & DAVID.STECKEL & SHAONE.MORRISONN & DAVID.STECKEL \\
\hline
\end{tabular}

\newpage
2010-2011: \\

\begin{tabular}{lccc}
\hline
Team & MVP Offense & MVP Defense & MVP Total \\
\hline
ANA & BOBBY.RYAN & GEORGE.PARROS & BOBBY.RYAN \\
ATL & DUSTIN.BYFUGLIEN & DUSTIN.BYFUGLIEN & DUSTIN.BYFUGLIEN \\
BOS & NATHAN.HORTON & JOHNNY.BOYCHUK & NATHAN.HORTON \\
BUF & DREW.STAFFORD & PAUL.GAUSTAD & PAUL.GAUSTAD \\
CAR & JEFF.SKINNER & BRANDON.SUTTER & BRANDON.SUTTER \\
CBJ & RICK.NASH & NIKITA.FILATOV & RICK.NASH \\
CGY & ALEX.TANGUAY & CORY.SARICH & JAROME.IGINLA \\
CHI & JONATHAN.TOEWS & BRIAN.CAMPBELL & BRIAN.CAMPBELL \\
COL & MATT.DUCHENE & DANIEL.WINNIK & DANIEL.WINNIK \\
DAL & BRAD.RICHARDS & JEFF.WOYWITKA & BRAD.RICHARDS \\
DET & BRIAN.RAFALSKI & TOMAS.HOLMSTROM & BRIAN.RAFALSKI \\
EDM & ALES.HEMSKY & THEO.PECKHAM & ALES.HEMSKY \\
FLA & CORY.STILLMAN & MIKE.WEAVER & MIKE.WEAVER \\
L.A & DREW.DOUGHTY & ALEC.MARTINEZ & DREW.DOUGHTY \\
MIN & BRENT.BURNS & GREG.ZANON & BRENT.BURNS \\
MTL & TOMAS.PLEKANEC & ROMAN.HAMRLIK & TOMAS.PLEKANEC \\
N.J & PATRIK.ELIAS & MARK.FAYNE & PATRIK.ELIAS \\
NSH & SIARHEI.KASTSITSYN & DAVID.LEGWAND & DAVID.LEGWAND \\
NYI & PA.PARENTEAU & ANDREW.MACDONALD & ANDREW.MACDONALD \\
NYR & RYAN.MCDONAGH & MICHAEL.SAUER & MICHAEL.SAUER \\
OTT & JASON.SPEZZA & RYAN.SHANNON & JASON.SPEZZA \\
PHI & JEFF.CARTER & ANDREAS.NODL & JEFF.CARTER \\
PHX & LAURI.KORPIKOSKI & SAMI.LEPISTO & SAMI.LEPISTO \\
PIT & SIDNEY.CROSBY & CHRIS.CONNER & SIDNEY.CROSBY \\
S.J & DEVIN.SETOGUCHI & KYLE.WELLWOOD & KYLE.WELLWOOD \\
STL & DAVID.BACKES & DAVID.BACKES & DAVID.BACKES \\
T.B & STEVEN.STAMKOS & BRETT.CLARK & STEVEN.STAMKOS \\
TOR & MIKHAIL.GRABOVSKI & FREDRIK.SJOSTROM & MIKHAIL.GRABOVSKI \\
VAN & HENRIK.SEDIN & KEVIN.BIEKSA & HENRIK.SEDIN \\
WSH & ALEXANDER.SEMIN & JOHN.CARLSON & ALEXANDER.SEMIN \\
\hline
\end{tabular}

%\newpage
%2010-2011: \\

\begin{tabular}{lccc}
\hline
Team & LVP Offense & LVP Defense & LVP Total \\
\hline
ANA & CAM.FOWLER & SAKU.KOIVU & CAM.FOWLER \\
ATL & JOHN.ODUYA & ANDREW.LADD & JOHN.ODUYA \\
BOS & BRAD.MARCHAND & MICHAEL.RYDER & BRAD.MARCHAND \\
BUF & SHAONE.MORRISONN & CRAIG.RIVET & CRAIG.RIVET \\
CAR & CHAD.LAROSE & ERIC.STAAL & CHAD.LAROSE \\
CBJ & NATHAN.PAETSCH & DERICK.BRASSARD & NATHAN.PAETSCH \\
CGY & RENE.BOURQUE & ALEX.TANGUAY & ALEX.TANGUAY \\
CHI & FERNANDO.PISANI & TOMAS.KOPECKY & TOMAS.KOPECKY \\
COL & RYAN.OBYRNE & KEVIN.PORTER & KEVIN.PORTER \\
DAL & TOM.WANDELL & BRENDEN.MORROW & TOM.WANDELL \\
DET & RUSLAN.SALEI & TODD.BERTUZZI & RUSLAN.SALEI \\
EDM & JASON.STRUDWICK & SAM.GAGNER & SAM.GAGNER \\
FLA & MIKE.WEAVER & DAVID.BOOTH & DAVID.BOOTH \\
L.A & TREVOR.LEWIS & JACK.JOHNSON & JACK.JOHNSON \\
MIN & ERIC.NYSTROM & MARTIN.HAVLAT & ERIC.NYSTROM \\
MTL & MAXIM.LAPIERRE & P.K..SUBBAN & MAXIM.LAPIERRE \\
N.J & DAVID.CLARKSON & ILYA.KOVALCHUK & DAVID.CLARKSON \\
NSH & NICK.SPALING & FRANCIS.BOUILLON & NICK.SPALING \\
NYI & BRUNO.GERVAIS & MILAN.JURCINA & BRUNO.GERVAIS \\
NYR & BRANDON.PRUST & MICHAEL.DEL ZOTTO & MICHAEL.DEL ZOTTO \\
OTT & ZACK.SMITH & BOBBY.BUTLER & BOBBY.BUTLER \\
PHI & DAN.CARCILLO & DARROLL.POWE & DAN.CARCILLO \\
PHX & DEREK.MORRIS & KYLE.TURRIS & DEREK.MORRIS \\
PIT & MAXIME.TALBOT & EVGENI.MALKIN & EVGENI.MALKIN \\
S.J & JAMIE.MCGINN & PATRICK.MARLEAU & PATRICK.MARLEAU \\
STL & B.J..CROMBEEN & JAY.MCCLEMENT & B.J..CROMBEEN \\
T.B & ADAM.HALL & DOMINIC.MOORE & ADAM.HALL \\
TOR & FREDRIK.SJOSTROM & TYLER.BOZAK & TYLER.BOZAK \\
VAN & TANNER.GLASS & MASON.RAYMOND & TANNER.GLASS \\
WSH & JEFF.SCHULTZ & JASON.CHIMERA & JASON.CHIMERA \\
\hline
\end{tabular}

\newpage
2011-2012: \\

\begin{tabular}{lccc}
\hline
Team & MVP Offense & MVP Defense & MVP Total \\
\hline
ANA & SAKU.KOIVU & SHELDON.BROOKBANK & SAKU.KOIVU \\
BOS & TYLER.SEGUIN & ADAM.MCQUAID & TYLER.SEGUIN \\
BUF & THOMAS.VANEK & ROBYN.REGEHR & THOMAS.VANEK \\
CAR & JAMIE.MCBAIN & PATRICK.DWYER & TIM.GLEASON \\
CBJ & VACLAV.PROSPAL & NATHAN.PAETSCH & VACLAV.PROSPAL \\
CGY & OLLI.JOKINEN & ROMAN.HORAK & ROMAN.HORAK \\
CHI & PATRICK.SHARP & NIKLAS.HJALMARSSON & PATRICK.SHARP \\
COL & GABRIEL.LANDESKOG & GABRIEL.LANDESKOG & GABRIEL.LANDESKOG \\
DAL & JAMIE.BENN & MARK.FISTRIC & JAMIE.BENN \\
DET & HENRIK.ZETTERBERG & TODD.BERTUZZI & IAN.WHITE \\
EDM & JORDAN.EBERLE & BEN.EAGER & JORDAN.EBERLE \\
FLA & STEPHEN.WEISS & JASON.GARRISON & STEPHEN.WEISS \\
L.A & JUSTIN.WILLIAMS & WILLIE.MITCHELL & JUSTIN.WILLIAMS \\
MIN & DANY.HEATLEY & NICK.SCHULTZ & NICK.SCHULTZ \\
MTL & ERIK.COLE & JOSH.GORGES & ERIK.COLE \\
N.J & PETR.SYKORA & JACOB.JOSEFSON & JACOB.JOSEFSON \\
NSH & MIKE.FISHER & BLAKE.GEOFFRION & BLAKE.GEOFFRION \\
NYI & JOHN.TAVARES & MATT.MARTIN & JOHN.TAVARES \\
NYR & MARIAN.GABORIK & RYAN.MCDONAGH & MARIAN.GABORIK \\
OTT & MARCUS.FOLIGNO & KYLE.TURRIS & MARCUS.FOLIGNO \\
PHI & SCOTT.HARTNELL & SEAN.COUTURIER & SCOTT.HARTNELL \\
PHX & RAY.WHITNEY & ADRIAN.AUCOIN & ADRIAN.AUCOIN \\
PIT & SIDNEY.CROSBY & DERYK.ENGELLAND & SIDNEY.CROSBY \\
S.J & PATRICK.MARLEAU & DOUGLAS.MURRAY & PATRICK.MARLEAU \\
STL & DAVID.PERRON & VLADIMIR.SOBOTKA & DAVID.PERRON \\
T.B & STEVEN.STAMKOS & BRETT.CONNOLLY & STEVEN.STAMKOS \\
TOR & JOFFREY.LUPUL & CARL.GUNNARSSON & JOFFREY.LUPUL \\
VAN & HENRIK.SEDIN & CHRISTOPHER.TANEV & HENRIK.SEDIN \\
WPG & BLAKE.WHEELER & JOHN.ODUYA & BLAKE.WHEELER \\
WSH & ALEXANDER.SEMIN & KARL.ALZNER & ALEXANDER.SEMIN \\
\hline
\end{tabular}

%\newpage
%2011-2012: \\

\begin{tabular}{lccc}
\hline
Team & LVP Offense & LVP Defense & LVP Total \\
\hline
ANA & JASON.BLAKE & CAM.FOWLER & JASON.BLAKE \\
BOS & SHAWN.THORNTON & DAVID.KREJCI & SHAWN.THORNTON \\
BUF & ROBYN.REGEHR & JASON.POMINVILLE & ROBYN.REGEHR \\
CAR & PATRICK.DWYER & ERIC.STAAL & ERIC.STAAL \\
CBJ & JOHN.MOORE & JEFF.CARTER & JOHN.MOORE \\
CGY & BLAKE.COMEAU & OLLI.JOKINEN & TIM.JACKMAN \\
CHI & MICHAEL.FROLIK & ANDREW.BRUNETTE & MICHAEL.FROLIK \\
COL & JAN.HEJDA & KYLE.QUINCEY & JAN.HEJDA \\
DAL & VERNON.FIDDLER & ADAM.PARDY & VERNON.FIDDLER \\
DET & TOMAS.HOLMSTROM & HENRIK.ZETTERBERG & TOMAS.HOLMSTROM \\
EDM & MAGNUS.PAAJARVI & JORDAN.EBERLE & MAGNUS.PAAJARVI \\
FLA & BRIAN.CAMPBELL & TOMAS.KOPECKY & BRIAN.CAMPBELL \\
L.A & TREVOR.LEWIS & ANZE.KOPITAR & TREVOR.LEWIS \\
MIN & NICK.SCHULTZ & DEVIN.SETOGUCHI & DEVIN.SETOGUCHI \\
MTL & MATHIEU.DARCHE & ALEXEI.EMELIN & ALEXEI.EMELIN \\
N.J & RYAN.CARTER & ILYA.KOVALCHUK & RYAN.CARTER \\
NSH & KEVIN.KLEIN & FRANCIS.BOUILLON & KEVIN.KLEIN \\
NYI & NINO.NIEDERREITER & MILAN.JURCINA & NINO.NIEDERREITER \\
NYR & MARC.STAAL & BRAD.RICHARDS & BRAD.RICHARDS \\
OTT & JARED.COWEN & STEPHANE.DA COSTA & JARED.COWEN \\
PHI & MATT.CARLE & HARRISON.ZOLNIERCZYK & MATT.CARLE \\
PHX & DEREK.MORRIS & DEREK.MORRIS & DEREK.MORRIS \\
PIT & JOE.VITALE & STEVE.SULLIVAN & JOE.VITALE \\
S.J & DOUGLAS.MURRAY & JAMIE.MCGINN & JAMIE.MCGINN \\
STL & SCOTT.NICHOL & ANDY.MCDONALD & SCOTT.NICHOL \\
T.B & DOMINIC.MOORE & STEVEN.STAMKOS & DOMINIC.MOORE \\
TOR & DAVID.STECKEL & JOFFREY.LUPUL & DAVID.STECKEL \\
VAN & DALE.WEISE & ALEXANDER.EDLER & DALE.WEISE \\
WPG & TANNER.GLASS & DUSTIN.BYFUGLIEN & TANNER.GLASS \\
WSH & MICHAEL.KNUBLE & TROY.BROUWER & MICHAEL.KNUBLE \\
\hline
\end{tabular}

\section{Exceptional Player Pairs, Overall}
\small
These ratings represent the total increase or decrease in team scoring rates if these two players play together, rather than separately. (We do not claim, for example, that the net impact of playing Crosby and Malkin is extremely negative, since their individual abilities are each positive; merely that their partnership leads to worse results than when these players play separately.)
\tiny
\begin{tabular}{lcccccc}
\hline
Rank & Player 1 & & Player 2 & & Total Time (s) & Rating \\
1 & BRAD.BOYES & R & JAY.MCCLEMENT & C & 35466 & 0.393 \\
2 & MATT.CARLE & D & ANDREJ.MESZAROS & D & 41011 & 0.314 \\
3 & PATRICE.BERGERON & C & BRAD.MARCHAND & C & 85678 & 0.31 \\
4 & JUSSI.JOKINEN & L & JEFF.SKINNER & C & 46196 & 0.287 \\
5 & KRIS.LETANG & D & PAUL.MARTIN & D & 40034 & 0.275 \\
6 & MICHAL.HANDZUS & C & WAYNE.SIMMONDS & R & 96815 & 0.265 \\
7 & TOM.GILBERT & D & RYAN.WHITNEY & D & 32378 & 0.247 \\
8 & BARRET.JACKMAN & D & KEVIN.SHATTENKIRK & D & 59537 & 0.24 \\
9 & JASON.BLAKE & L & DOMINIC.MOORE & C & 36983 & 0.231 \\
10 & PASCAL.DUPUIS & L & JORDAN.STAAL & C & 48978 & 0.225 \\
11 & KEITH.BALLARD & D & NICHOLAS.BOYNTON & D & 51300 & 0.212 \\
12 & ALEX.FROLOV & L & PATRICK.OSULLIVAN & C & 31117 & 0.211 \\
13 & VALTTERI.FILPPULA & C & JIRI.HUDLER & C & 115005 & 0.198 \\
14 & MATT.CARLE & D & CHRIS.PRONGER & D & 112174 & 0.197 \\
15 & MILAN.HEJDUK & R & WOJTEK.WOLSKI & L & 46147 & 0.191 \\
16 & KEVIN.BIEKSA & D & DAN.HAMHUIS & D & 102515 & 0.191 \\
17 & MARTIN.HANZAL & C & RADIM.VRBATA & R & 157171 & 0.188 \\
18 & KARL.ALZNER & D & JOHN.CARLSON & D & 120008 & 0.187 \\
19 & ALES.HEMSKY & R & DUSTIN.PENNER & R & 90640 & 0.186 \\
20 & MATT.GREENE & D & SEAN.ODONNELL & D & 42201 & 0.183 \\
21 & VERNON.FIDDLER & C & LEE.STEMPNIAK & R & 42605 & 0.181 \\
22 & CRAIG.CONROY & C & CURTIS.GLENCROSS & L & 44530 & 0.178 \\
23 & BRIAN.CAMPBELL & D & JASON.GARRISON & D & 62456 & 0.177 \\
24 & NICKLAS.LIDSTROM & D & BRIAN.RAFALSKI & D & 173306 & 0.177 \\
25 & ZACH.PARISE & L & TRAVIS.ZAJAC & C & 150049 & 0.176 \\
26 & MATT.GREENE & D & ALEC.MARTINEZ & D & 57029 & 0.175 \\
27 & TROY.BROUWER & R & PATRICK.SHARP & R & 38506 & 0.174 \\
28 & JIRI.HUDLER & C & HENRIK.ZETTERBERG & L & 63952 & 0.172 \\
29 & BRETT.CLARK & D & VICTOR.HEDMAN & D & 53748 & 0.169 \\
30 & SHANE.DOAN & R & STEVE.REINPRECHT & C & 51277 & 0.164 \\
31 & TONI.LYDMAN & D & LUBOMIR.VISNOVSKY & D & 88426 & 0.161 \\
32 & JOHNNY.BOYCHUK & D & ZDENO.CHARA & D & 92395 & 0.155 \\
33 & DREW.DOUGHTY & D & ROB.SCUDERI & D & 125662 & 0.154 \\
34 & ALEX.BURROWS & L & HENRIK.SEDIN & C & 173247 & 0.15 \\
35 & PATRICK.DWYER & R & BRANDON.SUTTER & C & 80414 & 0.148 \\
36 & PATRIK.ELIAS & L & BRIAN.ROLSTON & R & 57073 & 0.147 \\
37 & PAUL.MARTIN & D & DAVID.ODUYA & D & 85861 & 0.147 \\
38 & BRUNO.GERVAIS & D & MARK.STREIT & D & 66853 & 0.138 \\
39 & JOSH.GORGES & D & P.K..SUBBAN & D & 67710 & 0.138 \\
40 & ZDENO.CHARA & D & DENNIS.WIDEMAN & D & 82733 & 0.138 \\
41 & MATT.DUCHENE & C & MILAN.HEJDUK & R & 87232 & 0.133 \\
42 & FILIP.KUBA & D & PAUL.RANGER & D & 46711 & 0.131 \\
43 & VERNON.FIDDLER & C & TAYLOR.PYATT & L & 54765 & 0.124 \\
44 & CARLO.COLAIACOVO & D & ALEX.PIETRANGELO & D & 73311 & 0.123 \\
45 & MILAN.JURCINA & D & JEFF.SCHULTZ & D & 31131 & 0.12 \\
46 & BRAD.MARCHAND & C & GEORGE.PARROS & R & 32479 & 0.119 \\
47 & WILLIE.MITCHELL & D & SLAVA.VOYNOV & D & 37854 & 0.116 \\
48 & RYAN.CALLAHAN & R & CHRIS.DRURY & C & 72179 & 0.116 \\
49 & ROB.BLAKE & D & MARC-EDOUARD.VLASIC & D & 97836 & 0.109 \\
50 & TRAVIS.HAMONIC & D & ANDREW.MACDONALD & D & 107992 & 0.101 \\
\hline
201 & MICHAEL.CAMMALLERI & L & ANZE.KOPITAR & C & 29798 & -0.123 \\
202 & BRAD.BOYES & R & KEITH.TKACHUK & C & 68156 & -0.125 \\
203 & MILAN.HEJDUK & R & RYAN.SMYTH & L & 74613 & -0.125 \\
204 & DERICK.BRASSARD & C & RICK.NASH & L & 78314 & -0.125 \\
205 & DAVID.BACKES & R & ANDY.MCDONALD & C & 84713 & -0.128 \\
206 & MILAN.MICHALEK & L & JASON.SPEZZA & C & 104458 & -0.131 \\
207 & VYACHESLAV.KOZLOV & L & TODD.WHITE & C & 54517 & -0.132 \\
208 & FRANCOIS.BEAUCHEMIN & D & CAM.FOWLER & D & 74655 & -0.138 \\
209 & JAROME.IGINLA & R & OLLI.JOKINEN & C & 113358 & -0.145 \\
210 & TYLER.BOZAK & C & PHIL.KESSEL & R & 131533 & -0.148 \\
211 & MICHAEL.DEL ZOTTO & D & DAN.GIRARDI & D & 54129 & -0.157 \\
212 & MICHAEL.CAMMALLERI & L & JAROME.IGINLA & R & 52981 & -0.177 \\
213 & NIK.ANTROPOV & C & EVANDER.KANE & L & 39117 & -0.197 \\
214 & PAUL.STASTNY & C & CHRIS.STEWART & R & 65308 & -0.197 \\
215 & MARTIN.ST LOUIS & R & STEVEN.STAMKOS & C & 173577 & -0.206 \\
216 & KYLE.OKPOSO & R & JOHN.TAVARES & C & 60880 & -0.211 \\
217 & ZACH.BOGOSIAN & D & JOHN.ODUYA & D & 57215 & -0.235 \\
218 & DAVID.BOOTH & L & MICHAEL.SANTORELLI & C & 34158 & -0.241 \\
219 & ALEX.FROLOV & L & ANZE.KOPITAR & C & 45982 & -0.269 \\
220 & SIDNEY.CROSBY & C & EVGENI.MALKIN & C & 69217 & -0.283 \\
221 & ILYA.KOVALCHUK & L & TODD.WHITE & C & 70421 & -0.545 \\
\hline
\end{tabular}

\end{document}